# Software-testing education: A systematic literature mapping

Vahid Garousi
Queen's University Belfast
Northern Ireland, UK
v.garousi@qub.ac.uk
ORCID: 0000-0001-6590-7576

Austen Rainer
Queen's University Belfast
Northern Ireland, UK
a.rainer@qub.ac.uk
ORCID: 0000-0001-8868-263X

Per Lauvås jr
Kristiania University College
Oslo, Norway
per.lauvas@kristiania.no
ORCID: 0000-0003-1080-980X

Andrea Arcuri
Kristiania University College
Oslo, Norway
andrea.arcuri@kristiania.no
ORCID: 0000-0003-0799-2930

**Abstract**:

*Context*: With the rising complexity and scale of software systems, there is an ever-increasing demand for sophisticated and cost-effective software testing. To meet such a demand, there is a need for a highly-skilled software testing work-force (test engineers) in the industry. To address that need, many university educators worldwide have included software-testing education in their software engineering (SE) or computer science (CS) programs. Many papers have been published in the last three decades (as early as 1992) to share experience from such undertakings.

*Objective*: Our objective in this paper is to summarize the body of experience and knowledge in the area of software-testing education to benefit the readers (both educators and researchers) in designing and delivering software testing courses in university settings, and to also conduct further education research in this area.

*Method*: To address the above need, we conducted a systematic literature mapping (SLM) to synthesize what the community of educators have published on this topic. After compiling a candidate pool of 307 papers, and applying a set of inclusion/exclusion criteria, our final pool included 204 papers published between 1992 and 2019.

*Results*: The topic of software-testing education is becoming more active, as we can see by the increasing number of papers. Many pedagogical approaches (how to best teach testing), course-ware, and specific tools for testing education have been proposed. Many challenges in testing education and insights on how to overcome those challenges have been proposed.

*Conclusion*: This paper provides educators and researchers with a classification of existing studies within software-testing education. We further synthesize challenges and insights reported when teaching software testing. The paper also provides a reference ("index") to the vast body of knowledge and experience on teaching software testing. Our mapping study aims to help educators and researchers to identify the best practices in this area to effectively plan and deliver their software testing courses, or to conduct further education-research in this important area.

## TABLE OF CONTENTS

## 1 INTRODUCTION

*"Software is eating the world"* [1]. In other words, software systems have penetrated almost all industries and all aspects of our personal and professional lives. Furthermore, many industrial sources are reporting that software systems are getting increasingly complex [2-4]. With the increasing complexity and scale of software systems, there is an ever-increasing demand for sophisticated and cost-effective software quality assurance. Software testing is a fundamental, and also the most widespread, activity to assure the quality of software systems. A 2013 study by Cambridge University [5] reported that the global cost of locating and removing defects from software systems has risen to $312 billion annually, and removing defects comprises, on average, about half of the development costs of a typical software project.

To meet the ever-growing demand for cost-effective software testing, there is the increasing need for a highly-skilled software testing work-force in the industry. But in the non-peer-reviewed literature (grey-literature) such as online blogs and articles, many industrial sources are also reporting the *shortage* of software testers, e.g., [6-8]. Furthermore, in the context of software testing talent shortage, it is often recruiting "quality" software testers (i.e., with the right skillset), not just recruiting "quantity" (number of people available in the job market) [9] that is the challenge.

To address the above needs and to train more highly-skilled software testers, many software engineering (SE) and computer science (CS) university programs worldwide have started to include software testing in their education curricula. This is achieved by either including distinct and separate software testing courses, or alternatively blending (integrating) software testing concepts into programming or other courses [10]. To share their experience from such efforts, university educators write and publish papers about their software-testing education activities. Papers are written in this area to discuss and present more effective pedagogical approaches (e.g., how to better teach testing), new course proposals (e.g., course designs), and specific tools for testing education. Also, based on the educators' experience, many papers have identified a large number of challenges which educators and students could face when teaching and learning about testing, as well as insights into how to overcome those challenges.

Given such a large body of experience and knowledge in the area of software-testing education, there is the need for a systematic review in this area, since it is often not possible for the individual educator to study all the papers and to synthesize all the evidence presented. For example, for a new educator who wants to teach a (new) course in software testing (in her/his university), it would be very valuable to know, before teaching, about the challenges faced by educators when teaching testing and also about insights on how to overcome those challenges. Thus, it would be useful to synthesize and summarize reported experience and evidence, as well as research topics and research questions (RQs) in this area. Our objective and goal in this paper are to provide such a synthesis and summary.

To address the above goal, we conducted a systematic literature mapping (SLM) to synthesize what the community of educators has published on this topic. Based on a systematic SLM process, we systematically select a pool of 204 papers, and by investigating nine RQs, we categorize and analyze various aspects of the subject under study.

The contributions of this review paper are three-fold:

- The classification of the studies in this area, performed through a systematic mapping, via RQs 1-8
- The synthesis of challenges faced during testing education (via RQ 9.1) together with the synthesis of insights (recommendations) for testing education (via RQ 9.2).
- The development of an index (repository) of the studies in this area, which is accessible in an online Google spreadsheet (goo.gl/DcEpMv)

The remainder of this paper is structured as follows. Section 2 provides background and related work. Section 3 describes the research method, and then the design and execution phases of the SLM. Section 4 presents the results of the literature review. Section 5 summarizes the findings and discusses the lessons learned. Finally, in Section 6, we draw conclusions, and suggest areas for future research.

## 2 Background and Related Work

In this section, we provide a brief overview on the state of software-testing education in universities versus software testing training in industry. We then briefly review related work, i.e., existing survey (review) papers in the areas of software engineering and software-testing education.

### 2.1 Software-testing education in universities versus training in industry

In addition to software-testing education in universities, there is also high demand for software-testing training in industry. Industry training is often provided through certification schemes, such as those provided by the International Software Testing Qualifications Board (ISTQB) (www.istqb.org), an organization that has national branches in more than 120 countries worldwide. According to its website, "*As of December 2017, ISTQB has issued more than 570,000 certifications in over 120 countries world-wide*".

To more clearly understand, characterize and distinguish software-testing education in universities from training in industry, we model the concepts as a context diagram, as shown in Figure 1. To clarify the focus of this SLM paper, we have highlighted our focus with a grey background in Figure 1. On the left-hand side of Figure 1 are the higher-education institutions that train students and produce graduates. We further distinguish the SE, CS and IT degrees from non-SE/CS/IT degrees.

Educators in universities may decide to include or not include testing in their SE, CS, or IT curricula. As a result, on a world-wide scale, we have graduates of SE, CS, and IT degrees with varying levels of opportunity to learn software testing during their university studies. These graduates look for positions in industry and are employed as SE professionals. We also recognize that, in the software industry, many graduates of *non*-SE/CS/IT degrees (e.g., math, or business) also work in software testing positions. For example, in a survey of software testing practices in Canada in 2013 [11], based on a respondent population of 246 practitioners, 92 respondents (37.3% of all respondents) reported having non-SE/CS/IT degrees, e.g., business, MBA, industrial engineering, mathematics, English and sports administration.

To do a better job in software testing, and/or to find better positions, university graduates and practicing SE professionals sometimes also self-learn (self-train) [12, 13] in software testing by learning from books or online resources, or they attend industrial training and achieve certification in software testing, e.g., those provided by ISTQB.

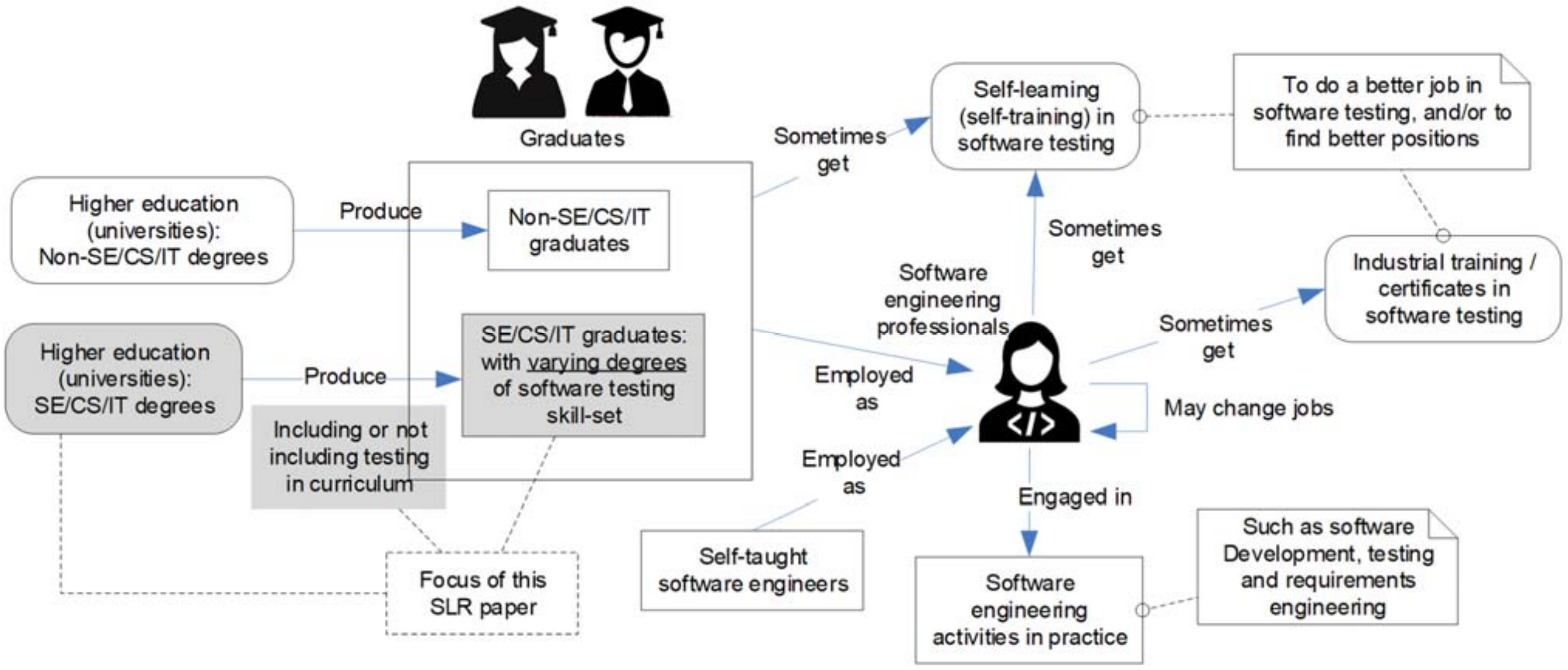

**Figure 1-A context diagram modelling the relationship of software-testing education in universities versus training in industry**

### 2.2 RELATED WORKS: SECONDARY STUDIES IN SOFTWARE ENGINEERING EDUCATION

Many systematic review papers have been reported in the general area of SE education, and CS education. Also, review papers have been reported on educational aspects for specific sub-areas of SE, e.g., testing, or requirements engineering. Based on a (non-exhaustive) literature search, we present in Table 1 a list of those studies. For each review paper, we also provide the year of publication, number of papers reviewed in the review and type of review: regular survey, Systematic Mapping Study (SMS) which is also called Systematic Literature Mapping (SLM) [14], or Systematic Literature Review (SLR). Papers are sorted by year of publication, for each category.

Note that, because our focus is on software-testing education, the search for review papers in this area was done more carefully, and thus we believe the five papers shown in the software-testing category in Table 1 are all that have been published so far on this topic. To keep our discussion focused, we discuss next only those five related review papers [15-19] and how this current SLM differs from them. To also help us differentiate our SLM, we list in Table 2 the RQs raised and studied in the five review papers. We also summarize in Table 2 how the RQs of each previous review study relate to the RQs in this SLM.

A survey of evidence for Test-Driven Development (TDD) in academia was reported in [15]. The work was a regular survey (not necessarily systematic). By reviewing 18 papers on the topic, it presents the benefits of incorporating TDD in testing education, "worries" (challenges) of doing so, and popular frameworks for that purpose.

A SLM on software-testing education (paper written in Portuguese) was reported in [16]. It reviewed 25 papers on the topic, published as of 2015. The study identified the approaches of teaching software testing, as well as how to develop and evaluate them.

Another SLR [17] synthesized the challenges of integrating software testing into introductory programming courses, as reported by 158 papers, which included: (1) Determining how programming and testing should be connected and delivered together; (2) Dealing with students who do not appreciate the value of software testing; (3) Determining how the testing activity should be conducted in programming assignments; (4) How to help students become better testers; and (5) Choosing appropriate tools. The study also discussed possible solutions to the challenges as addressed in the literature.

Recent trends in software-testing education were studied via a SLR in [18]. The SLR analyzed and reviewed 30 papers that were published between 2013 and 2017. The review pointed out recent trends such as the use of gamification to make the software testing more interesting and less tedious for students. Two of the current authors were involved in that SLR.

The SLM reported in the current paper is a substantial extension to the SLR [18], since: (1) we have extended the pool of papers under study from 30 to 204 papers; (2) we have also extended our analysis from only three RQs in [18] to nine RQs (discussed in Section 3.1).

A recent SLM was published in 2019 [19] which explored integration of software testing in introductory programming courses. By populating a large pool of 293 papers, it provided a mapping on two RQs (as shown in Table 2).

**Table 1: An indicative list of survey papers in the area of software engineering education**

| Area / category | Year | Reference of the review paper | Num. of papers reviewed | Type of review | | |
|---|---|---|---|---|---|---|
| | | | | Regular survey | SLM | SLR |
| Software-testing education | 2008 | [15] | 18 | x | | |
| | 2015 | [16] | 25 | | x | |
| | 2017 | [17] | 158 | | | x |
| | 2018 | [18] | 30 | | | x |
| | 2019 | [19] | 293 | | x | |
| | **2019** | **This SLM paper** | **204** | | **x** | |
| Requirements-engineering education | 2015 | [20] | 79 | | x | |
| Software process education | 2015 | [21] | 33 | | x | |
| Other areas of software engineering education | 2011 | [22] | 36 | | | x |
| | 2012 | [23] | 70 | | x | |
| | 2013 | [24] | 53 | | x | |
| | 2014 | [25] | 173 | | x | |
| | 2018 | [26] | 127 | | x | |
| | 2018 | [27] | 156 | | x | |
| | 2019 | [28] | 34 | | | x |
| Automated assessment approaches of programming assignments | 2005 | [29] | 65 | x | | |
| | 2010 | [30] | 80 | | | x |
| Computer science education | 1977 | [31] | 200 | x | | |
| | 1988 | [32] | 17 | x | | |

We have summarized in Table 2 how the RQs of each previous review study relate to the RQs in this SLM. By comparing the RQs listed in Table 2, we can find out that the focus of our SLM (as indicated by our nine RQs) is wider than the focus of the previous secondary studies in this area, since those previous review studies have had between one and four RQs, each.

In terms of a study's *substantive* pool size, we believe our SLM to be the largest review. The 2019 SMS study [19] – henceforth referred to as the SMS2019 study – has reviewed more primary studies (293), however after examining the paper pool of the SMS2019 study, we conclude that the study [19] included, in its pool of papers, many papers which had not focused mainly on testing education, but were rather: papers presenting experience in *programming* education together with, in some cases, short discussions relating to testing, e.g., [33-37]; or papers focused on *automated assessment* systems for student-written programs (assignments), e.g., [29, 38-43], a topic which we believe is not directly about "teaching" software testing, and thus should not be included when doing a SLR on software-testing. By contrast, for our SLM, we only included papers with a clear focus on testing education. The SMS2019 study [19] also included other "secondary" studies in its pool, i.e., a "secondary" study [44] is a study of studies (papers) and is another term used to refer to survey/review studies. However, we question whether a secondary study should include other secondary studies in its review pool. If a study intends to review and synthesize secondary studies, it would then become a "tertiary" study [45-47].

To further clarify the position of our SLM in relation to previous review papers in this area, we provide a Venn-diagram visualization in Figure 2. As software testing can sometimes be taught as part of a programming course, our SLM reviews some publications that overlap with the teaching of programming. The figure indicates two previous studies - SMS2017 and the already-mentioned SMS2019 (see Table 1) - that specifically focus on programming and testing. Software testing also provides a mechanism for automated assessment of students' programming, so our SLM also includes some publications that overlap with automated assessment. There are many other areas to software engineering education, in addition to programming and automated assessment. We present two areas in Figure 2, as examples: requirements engineering and software process. Thus, we believe our SLM to be the most comprehensive SLR published to date on software-testing education.

**Table 2: The RQ raised and studied in the survey papers in software-testing education**

| Title of the paper | ID in Figure 2 | Reference | RQs<br>How the RQs of each previous review relate to our RQs (discussed in grey background) |
|---|---|---|---|
| A survey of evidence for test-driven development in academia | SMS2017 | [15] | No formal RQ, since it was a regular survey paper. But the survey extracted the following information from the primary studies:<br>• Type of student levels (junior, graduate, etc.)<br>• Number of subjects (students); Corresponds to one of the aspects covered in our RQ 6 (Context under study)<br>• Evidence for increase in productivity of students<br>• Evidence for increase in quality of programs |
| A systematic mapping on software-testing education *(Paper is in Portuguese)* | - | [16] | (note: since the paper is in Portuguese, we used the Google Translate to translate its RQ phrases into English)<br>• RQ1 - What are the types of approaches that have been used to aid teaching software testing?<br>o This RQ covers only one aspect of the paper contribution types (RQ 1) in our study.<br>• RQ2 - What are the phases of software testing that have been contemplated in teaching software testing?<br>o This RQ is similar to our RQ 5.1 (Type of test activities covered in the course).<br>• RQ3 - What technologies have been used in the development of the approaches identified in RQ1 and what are the target languages used to aid the teaching of software testing?<br>o This RQ is similar to our RQ 5.2.<br>• RQ4 - What are the evaluations that have been carried out for the validation of the approaches used to support the teaching of software testing?<br>o This RQ is similar to our RQ 3. |
| Challenges to integrate software testing into introductory programming courses | - | [17] | One RQ: What are the challenges faced to integrate testing practices into introductory programming courses?<br>This RQ has some similarity to our RQ 9.1 (Challenges in testing education) |
| Recent trends in software-testing education: a systematic literature review | - | [18] | • RQ1: What topics are addressed in the recent literature on software-testing education?<br>o This RQ is similar to our RQ 1.<br>• RQ2: What kind of contributions are present in papers published on software-testing education?<br>o This RQ is similar to our RQ 2.<br>• RQ3: What insight can be provided to lecturers that want to introduce software testing as part of their teaching?<br>o This RQ is similar to our RQ 9.2. |
| Software testing in introductory programming courses: a systematic mapping study | SMS2019 | [19] | • RQ1: Which topics have researchers investigated about software testing in introductory programming courses?<br>o This RQ is similar to our RQ 1.<br>• RQ2: What are the benefits and drawbacks about the integration of software testing into introductory programming courses?<br>o This RQ has some similarity to our RQ 9.1 (Challenges in testing education) and RQ 9.2 (Insights for testing education). |

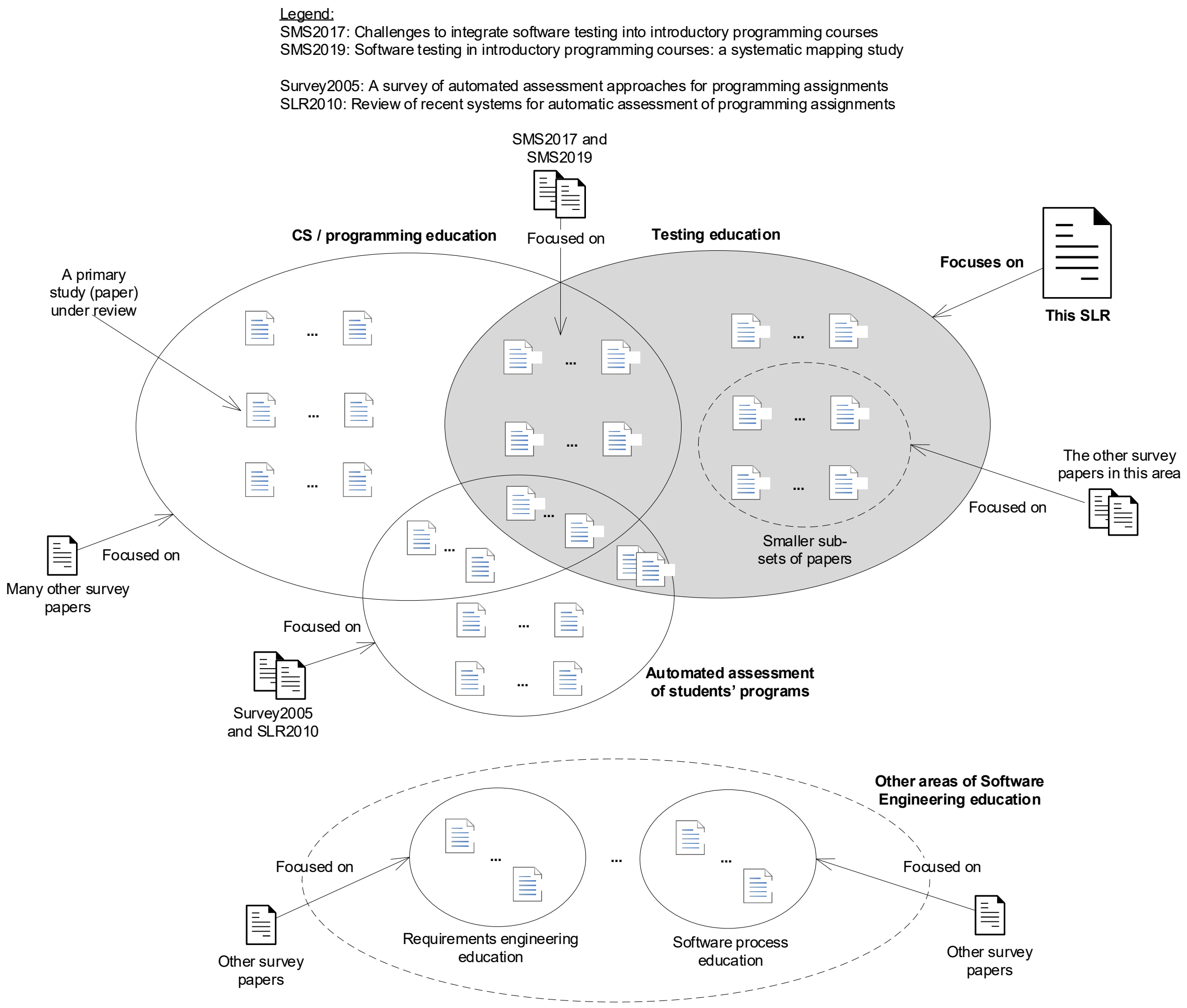

**Figure 2-A Venn diagram showing the scope of our SLM and other review studies w.r.t. topic areas assessed**

## 3 DESIGN AND EXECUTION OF THE SLM

Our research method in this paper is SLM. We present an overview to our SLM process, the research questions of the SLM, and then other aspects related to the design and execution of the SLM.

Using the well-known guidelines for conducting SMS and SLR studies in SE (e.g., [14, 48-50]) and also based on our past experience in SLR studies, e.g., [51-56], we developed our SLM process, as shown in Figure 3. We discuss the SLM planning and design phase (its goal and RQs) in the next section. We then present in Section 4 each of the follow-up phases of the process.

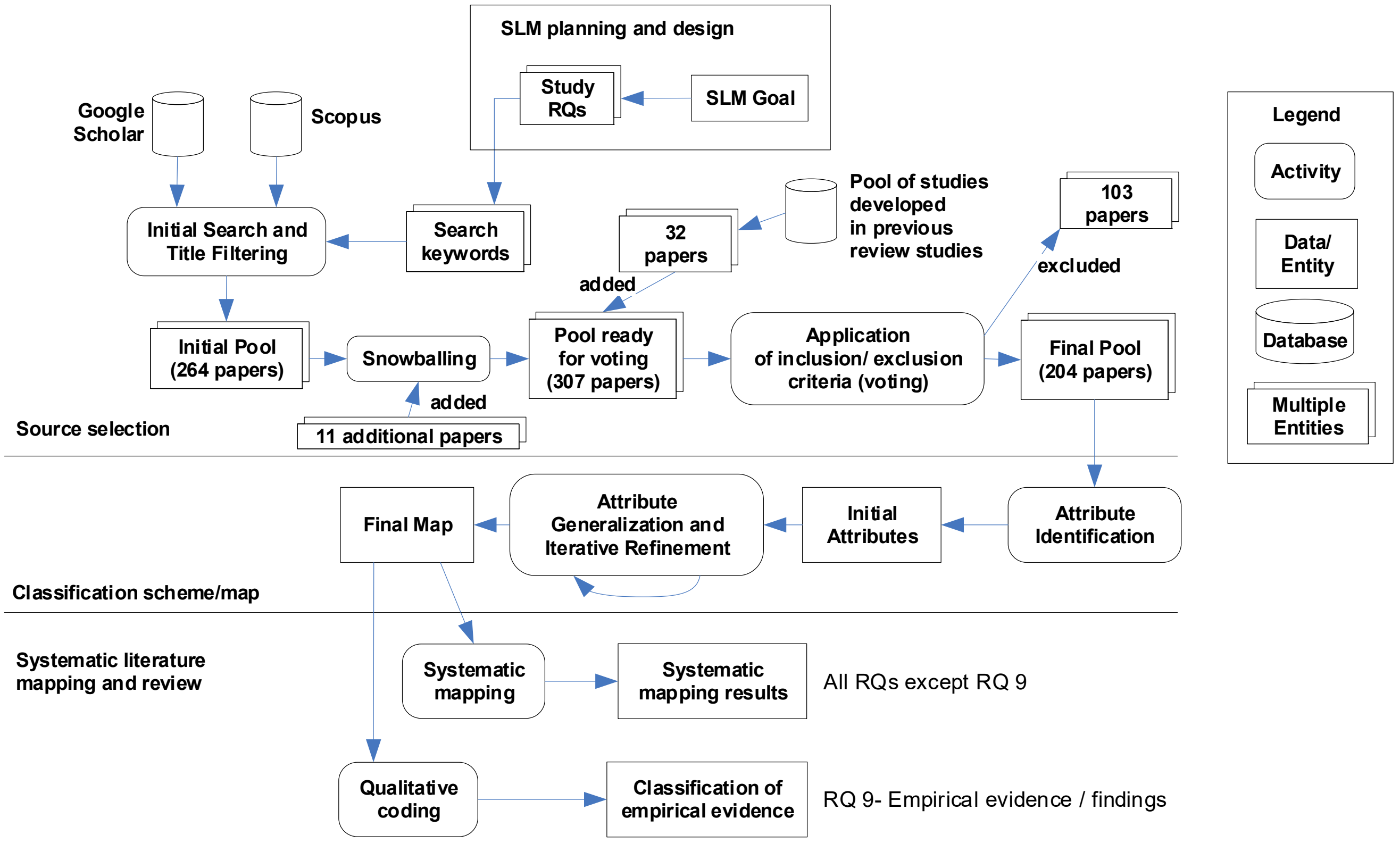

**Figure 3-An overview of SLM process (represented as a UML activity diagram)**

## 3.1 GOAL AND RESEARCH QUESTIONS

The goal of this study is to classify and summarize reported experience and evidence, as well as research topics and research questions, in the area of software testing education. By doing so, we seek to provide a holistic view to the body of knowledge on software testing education.

As discussed in Section 1, the need for conducting this SLM study was established by identifying the relevant stakeholders and their needs, as follows: (1) new and also experienced testing educators, and also (2) education researchers in this area, whose needs are as follows. For a new educator who wants to teach a (new) course in software testing (in her/his university), it would be valuable to know, before teaching, about the challenges faced by educators when teaching testing and also about insights on how to overcome those challenges. It would also be helpful for him/her whether certain tools / evidence have been reported in this area to support testing education. The authors have been involved in teaching testing for many years. They clearly remember when they started teaching, they were looking for experience and evidence shared by others in the literature. In many occasions, they had to find many papers and review them individually. Having a single holistic overview / classification (which is provided by this SLM) would have been helpful for such needs. The authors actually know several junior colleagues teaching courses in software testing and they have expressed interest in seeing results of our SLM study.

Furthermore, education researchers in this area need to have a clear view on the state of the art to be able to properly plan and conduct new research in the topic. Among many research aspects, an education researcher in this area would need to know the contribution types presented and research methods used in previous studies, e.g., experiments and empirical studies.

Based on the above goal, we raise the following research questions (RQs):

- **RQ 1- Classification of studies by contribution types**:
    - **RQ 1.1-**What contributions to software-testing education have been made by the papers in this area, and what are their frequencies?

- **RQ 1.2-**What 'clusters' of papers are there (if any) and how do they cluster? For example, during our reviews we have identified particular tools around which several papers are written, and particular sets of authors appear to write a collection of papers together, over time.
- **RQ 2- Classification of studies by research method types**: What research methods have been used in the papers and what are their frequencies of use? Some papers have presented proposals or demonstration of a course or a tool but share very little experience, while some other papers have shared more in-depth experience of the proposed courses, such as students' feedback. Some papers have gone further and have conducted systematic experiments or empirical studies by raising Research Questions (RQs). The rationale for this RQ is that the education researchers in this area who plan to design and conduct new empirical research in the topic would benefit from rigorous empirical studies which have already been published (to be identified by this RQ).
- **RQ 3- Data source (if any) used for the evaluation:** What types of data sources were used in the evaluations? For example, we found that some papers used survey data (gathered from students), while others analyzed the quality of student-written automated tests.
- **RQ 4- Research questions, or hypotheses, studied in the papers:** What research questions or hypotheses have been raised and studied? Knowing about the RQs studied in the papers could inspire readers (other researchers) to explore similar or other research directions in future.
- **RQ 5- Technical aspects of testing:** What technical aspects of testing (test activities) have been covered in testing education, as reported in the primary studies? Our rationale for this RQ is to assess the coverage of different technical testing topics in the courses, as discussed in the papers, e.g., test process planning, test-case design and test automation. We wondered whether certain test activity types (e.g., test-case design) had received more attention in terms of teaching coverage than others, while certain activity types (such as test automation) has received less attention: will the general landscape of research in this area (as analyzed by the SLM) support such propositions?
- **RQ 6- Scale of the educational setting under study:** What were the number of the course offerings, and also the number of students taking the course(s)? Our rationale was to get a measure of the scale/context of the software testing courses, used for evaluations in the papers. According to the empirical software engineering guidelines [57], one would expect that larger empirical contexts (e.g., having more students as subjects of empirical studies) would lead to better/stronger evidence in our subject matter.
- **RQ 7- Different approaches to testing education:** How many of the papers have presented a single testing course and how many have integrated testing across one or more programming courses?
- **RQ 8- Theories and theory-use in software-testing education:** What is the state of theories and theory-use in software-testing education? As it has been recognized in the broad literature of CS/SE education, e.g., [58-60], it is important to use and adapt theories from learning and education science in CS/SE education to increase research rigor in this area.
- **RQ 9- Empirical evidence collected and reported:** What types of empirical evidence have been collected and reported? We organize this RQ into two sub-RQs:
  - **RQ 9.1- Evidence-based challenges in testing education:** What challenges in testing education have been supported by empirical evidence?
  - **RQ 9.2-Evidence-based insights (recommendations) for testing education:** What insights (recommendations) for testing education have been supported by empirical evidence?

### 3.2 The search process: Selecting the source engines and search keywords

For all steps of this SLM, we have followed the common practices employed in the SLR and SLM studies in software engineering [14, 48-50]. To find and select the papers for the SLM, we used both Google Scholar and Scopus. These are both widely used in many previous SLR and SLM papers in software engineering, e.g., [61-64]. The reason that we used Scopus in addition to Google Scholar was that several sources have mentioned that: "*it [Google Scholar] should not be used alone for systematic review searches*" [65] as it may not find all papers.

All the authors did independent searches using the search strings. During this search phase, the authors already applied inclusion/exclusion criteria for selecting only those papers which explicitly addressed the study's topic. We show in Table 3 our search strings used in each search engine. For each case, we also display the number of records (papers) returned, number of papers added to the candidate pool, and number of papers not added (as they were already in the pool).

Furthermore, to ensure maximizing our chances of finding all the relevant papers, we identified two well-known focused venues in this topic: (1) the Conference on Software Engineering Education & Training (CSEE&T), and (2) the ACM SIGCSE

Technical Symposium on Computer Science Education, and searched in their proceedings directly (see the URLs in Table 3).

**Table 3: Search engines and search strings**

| | Sources | Search string | # of records returned | # of papers added to the candidate pool | # of papers not added (already in the pool, or immediately excluded) |
|---|---|---|---|---|---|
| **Search engines:** | scholar.google.com | Software testing education OR Teaching software testing | ~ 4,390,000 | 226 | N/A |
| | www.scopus.com | (TITLE-ABS-KEY (software testing education) OR TITLE-ABS-KEY (teaching software testing )) AND SRCTITLE ( software ) | 309 | 20 | 289 |
| **Focused venues in this topic:** | Conference on Software Engineering Education & Training (CSEE&T) https://ieeexplore.ieee.org/xpl/conhome.jsp?punumber=1000686 | testing | 131 | 5 | 126 |
| | ACM SIGCSE Technical Symposium on Computer Science Education https://dl.acm.org/event.cfm?id=RE175 | testing | 574 | 13 | 561 |
| | | | Total in the initial pool: | 264 | - |

These searches were conducted in January 2019. The data extraction from the primary studies and their classifications were conducted during the period of January-February 2019. As we wanted to include all available papers on software testing education regardless of when they were published, we did not restrict the papers' year of publication (e.g., only those published since a specific year, like for example since 2000).

While performing the searches with the selected keywords, we also applied title filtering. We wanted to ensure that we would add to our candidate paper pool only those papers that were obviously or potentially relevant papers, while at the same time we wanted to avoid wasting time analyzing non-relevant papers. It would be a waste of effort to add a clearly irrelevant paper to the candidate pool and then remove it soon after. Our first inclusion/exclusion criterion (discussed in Section 4.1.2) was used for this purpose (i.e., does the source focus on software-testing education?). For example, Figure 4 shows a screenshot of our search activity using Google Scholar in which obviously or potentially relevant papers are highlighted by red boxes. To ensure efficiency of our efforts, we only added such related studies to the candidate pool.

Google Scholar returned a very large number of results (papers) using the above keyword, at the time of our search phase (January 2019), i.e., more than 2 million papers. Analyzing all of them would simply not be viable. We needed a clear, unbiased "stopping" condition to determine how many papers should be considered. To cope with this issue, we utilized the relevance ranking of the search engine (Google's PageRank algorithm) to restrict the search space. Fortunately, the PageRank algorithm is very effective at ranking the relevant search results. Thus, we checked only the first $n$ pages (i.e., somewhat like a search "saturation" effect). We continued with further pages of results only if needed, e.g., when at least one result in the $n^{th}$ page was still relevant (for example, if at least one paper focused on software testing education). Similar heuristics have been reported in several other existing review studies, guidelines and experience papers [66-70]. At the end of our initial search and title filtering, our candidate pool had 307 papers (as shown in our SLM process in Figure 3).

There were high chances of duplications in the pool, as Scopus and Google Scholar databases are not independent. Therefore, we added a candidate paper to the paper pool only if it was not already in the candidate pool.

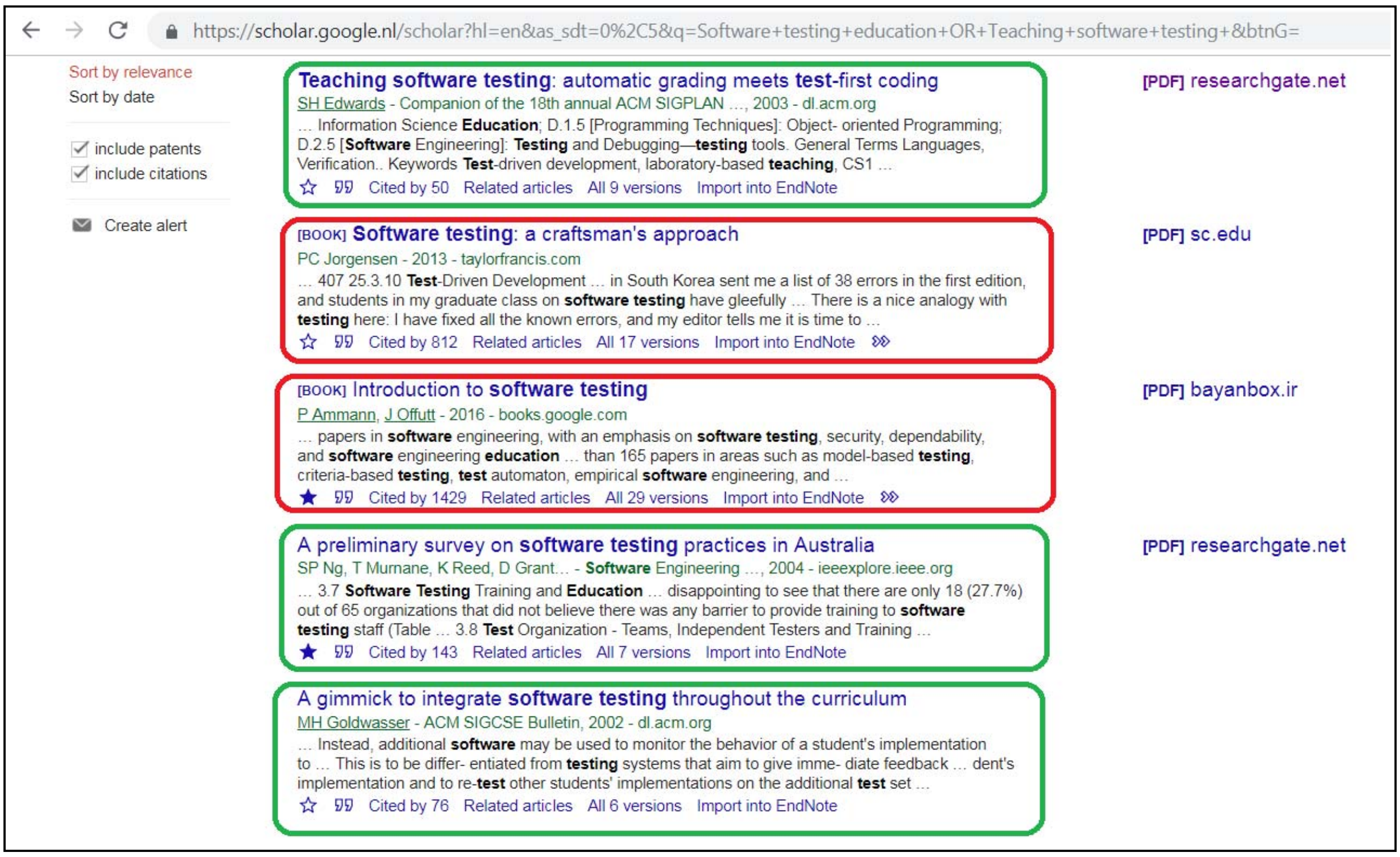

**Figure 4- A screenshot from the search activity using Google Scholar (directly- or potentially-relevant papers are highlighted by red boxes)**

We conducted forward and backward *snowballing* [48] on the set of papers already in the pool. The aim was to ensure to include all the relevant sources as much as possible, as recommended by systematic review guidelines. Snowballing, in this context, refers to using the reference list of a paper (backward snowballing) or the citations to the paper to identify additional papers (forward snowballing) [48]. Snowballing provided 11 additional papers. For example, we found [P14] and [P117] by backward snowballing of [P114]. Note that the citations in the form of [Pn] refer to the IDs of the primary studies (papers) reviewed in our study. They are available in an online archived document [71].

Referring to the process of how we populated the pool of papers, as shown in Figure 3, we should clarify the case of 32 papers which we added from the pool of papers of previously-published review studies. Of course, the previously-published review studies had reviewed considerably more than 32 papers. We went through their pools of papers and only added to our pool the candidate papers which were not *already* in the initial pool of 275 papers (=264 'original' papers + 11 'snowballed' papers).

After compiling an initial pool of 307 candidate papers, a systematic voting (as discussed next) was conducted among the authors, in which a set of defined inclusion/exclusion criteria were applied to derive the final pool of the primary studies.

We should add that, although any secondary study like ours seeks to apply comprehensive search processes and measures to ensure that we are "*not missing relevant [primary] studies*" [72], one can only minimize the chance of missing related primary studies, and there is no absolute guarantee that we find all papers. Indeed, there are papers outside the SE discipline that focus on the issue of effectiveness of search approaches, e.g., [72-74]. For example, Cooper et al. mention in [72] that: "*… comprehensive literature searching is implicitly linked to not missing relevant studies*". By following such recommendations, we designed and conducted our comprehensive search process, as discussed above.

### 3.3 Application of inclusion/exclusion criteria and voting

We carefully defined a set of three exclusion criteria to ensure including all the relevant papers and excluding the out-of-scope papers. Our exclusion criteria were as follows:

- *Exclusion criterion #1*: Does the paper focus on software-testing education in academia (universities)? Our scope was to exclude papers reporting software-testing training in industry.

- *Exclusion criterion* #2: Does the paper include a relatively sound evaluation, e.g., at least based on some teaching experience? We wanted to exclude purely opinion-based papers.
- *Exclusion criterion* #3: Is the paper in English and can its full-text be accessed on the internet?

For each of the exclusion criteria, we sought a binary answer to the question: either Yes (value=1) or No (value=0). Our voting approach was as follows. One of the researchers voted on all the candidate papers using the above criteria. The other researchers then peer reviewed all those votes. Disagreements were discussed until consensus was reached for all papers.

When we were assessing the candidate papers using the above exclusion criteria, we also carefully assessed the relationships (if any) among the papers, i.e., if they had similar topics, or were written by the same author(s). We created a log of such "inter-related" papers, to form "clusters" of papers. For example, we found that 10 papers have been published focusing on a specific web-based system for automated testing and grading of programming assignments (named *Web-Cat*), i.e., [P3, P65, P87, P90, P96, P130, P132, P200, P198, P204]. We also used the log of inter-related papers later to answer RQ 1.2 (concerning the clusters of similar papers in terms of topics), in Section 5.1.2.

We included only the papers which received 1's for all the three criteria, and excluded the rest. Application of the above criteria led to the exclusion of 103 papers. Details on the 103 excluded papers is provided with the study's online spreadsheet. Excluded papers were classified based on the exclusion reasons, as summarized in Table 4. 71 candidate papers were excluded for failing to satisfy the inclusion criteria #1. For example, paper [75] was an empirical study whose goal was to reveal effective and ineffective software testing behaviors of novice programmers. While the paper studied students participating in software testing, the paper's focus was not on software testing *education*. As a second example, paper [76] presents an undergraduate course on software bug detection tools and techniques. The paper covered topics such as symbolic execution, constraint analysis, and model checking, but not testing. Two sentences in the paper explicitly state, "*This course is not to be confused with a course on software testing*", and "*A separate course on software testing would complement this course*".

21 candidate papers were excluded for failing to satisfy exclusion criteria #2, e.g., [77, 78]. Paper [77] was a half-page document referring to a tutorial titled "*Adding software testing to programming assignments*" offered during a conference in 2005. Paper [78] was a "position paper". As our work is a secondary study (i.e., SLM) itself, we only included "primary" studies and did not include secondary (literature review) studies in our pool of candidate papers.

**Table 4: Statistics of excluded papers**

| Exclusion reason | # of excluded papers due to this reason |
|---|---|
| Criterion #1 (Does the paper focus on software-testing education?): Out of scope | 71 |
| Criterion #2 (Is the paper based on experience?): Excluding purely opinion and "position papers" and tutorials at conferences | 21 |
| Criterion #3 (Is the paper in English and can its full-text be accessed on the internet?) | 10 |
| A longer version of the paper is already included in the pool | 1 |

### 3.4 Final pool of the primary studies

As mentioned above, the references for the final pool of 204 papers can be found in an online archived document [71]. To provide transparency and enable replicability of our analysis, full details of the extracted data from all the papers are also available in an online Google spreadsheet (goo.gl/DcEpMv). Once we finalized the pool of papers, we first wanted to assess the growth of this field by the number of published papers each year. For this purpose, we depict in Figure 5 the annual number of papers (by their publication years).

As discussed in Section 4.1, since we searched for the papers in January 2019, the number of papers for 2019 is partial and thus low (only 2 papers). The earliest paper in this area was published in 1992, and was titled "*Assessing testing tools - in research and education*" [P37]. Research activity in this area peaked up in early 2000's and, since the year 2010, about 13-16 papers are published each year.

To put things in perspective, we compare the annual trend of papers with the trend data for five other software testing areas as reported by five other SLM/SLR studies: (1) an SLR on testing embedded software [79], (2) an SLR on web application testing [80], (3) an SLR on Graphical User Interface (GUI) testing [81], (4) a survey on mutation testing [82], and

(5) an SLR on software testability [1]. Note that, as we can see in Figure 5, the trend data for the other different SLRs end in different years, due to timeline differences in the execution and publication of those survey papers, e.g., the survey on mutation testing [82] was published in 2011 and thus only has the data until 2009. But still, the figure provides a comparative view of the growth of these six sub-areas of software testing.

As one can see in Figure 5, papers on testing embedded software and testability started a bit "earlier" (in the early 1980's) than software-testing education. This could be because it took a while for educators to be involved in software-testing education and then see the value in sharing experience on this topic. "Technical" areas such as testing embedded software and mutation testing seem to have more annual papers than testing education, which seems reasonable due to more research activity taking place on those topics.

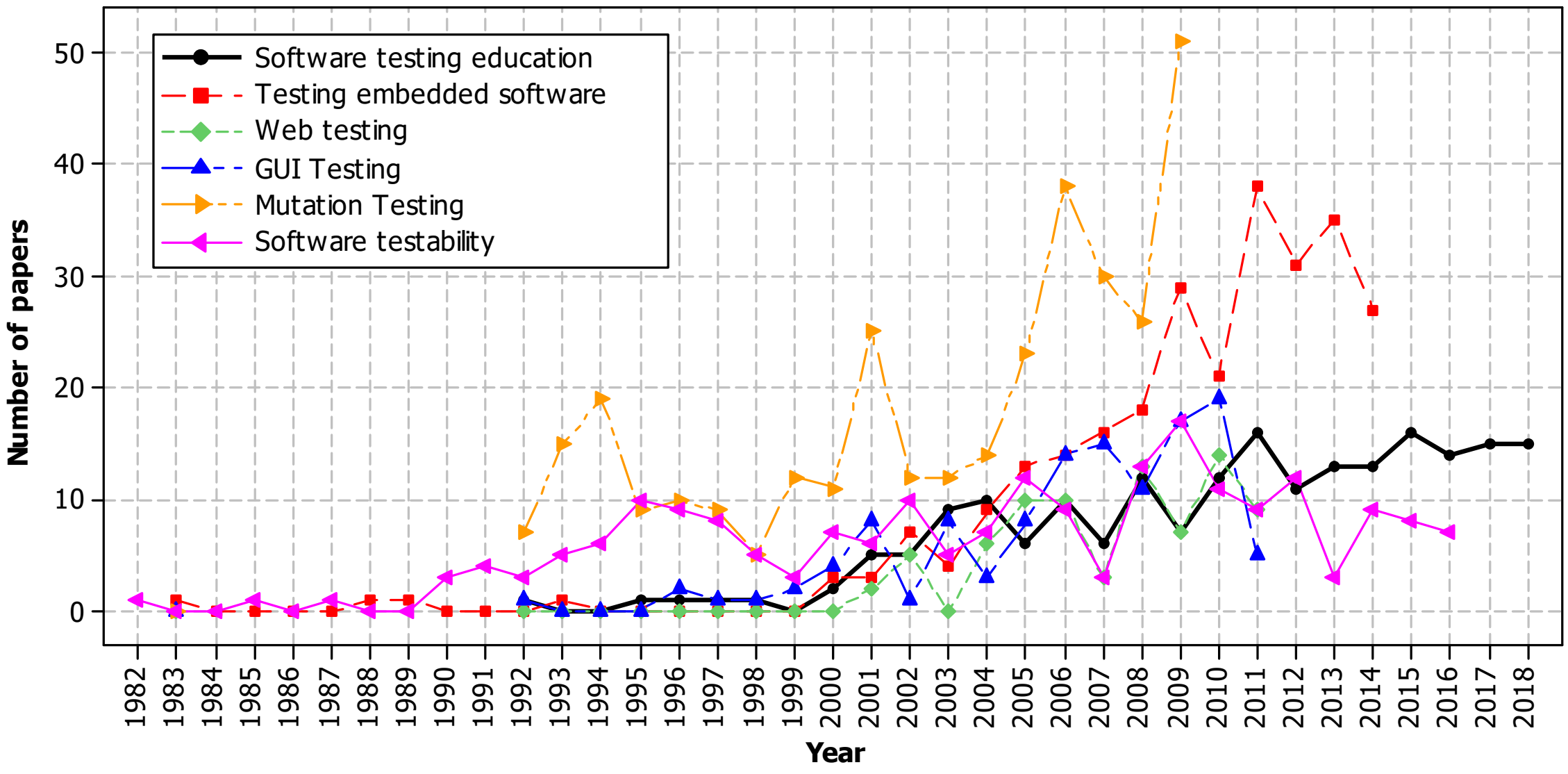

**Figure 5-Growth of the field (software-testing education) and comparing the growth with five other software testing areas**

### 3.5 Development of the systematic map

To answer each of our research questions, we developed a systematic map (also known as "classification scheme"). Then, we extracted the relevant data from the papers in our pool, and classified them using our systematic map. Next, we discuss how we developed such systematic map.

To develop our systematic map, we analyzed the studies in the pool and identified the initial list of relevant attributes. As shown in Figure 3, the final map was derived by using attribute generalization and iterative refinement, when necessary.

To facilitate further analyses, all relevant papers were recorded in an online spreadsheet. Our next goal was then to categorize these studies, in order to begin building a complete picture of the research area and to answer the study RQs. We refined these broad interests into a systematic map using an iterative approach.

Table 5 shows the final classification scheme that we developed after applying the process described above. In the table, column 2 is the list of RQs, column 3 is the corresponding attribute/aspect, and column 4 is the set of all possible values for the attribute. The fifth column indicates, for an attribute, whether multiple selections could be made. For example, for RQ 1.1 (contribution type), the corresponding value in the last column is "Multiple", indicating that one study can be classified with more than one contribution type (e.g., method, tool). In contrast, for RQ 2 (research-method type), the corresponding value in the last column is "Single", indicating that one paper can be classified under only one research-method type.

Among the research types, the least rigorous type is "Proposal" in which a given paper only presents a course proposal with no or very little educational experience of implementing that course professionally. In levels 2 and 3 of research method types, we have: (2)-Experience / Informal evaluation and (3)-Experiment / empirical study, with increasing levels of maturity as mentioned by the wordings. We also observed that a few papers had conducted reviews of practices when teaching testing, e.g., [P15] which presented an opinion survey on graduates' curriculum-based knowledge gaps in software testing.

The last column of Table 5 shows our approach to answer each of the RQs or sub-RQs based on the extracted data. As we can see, six of the RQs are addressed by classification. For two cases, the RQs will be addressed by simply merging all the extracted data: RQ 4 and RQ 7.2. To answer RQ 6, concerning the scale of the educational setting under study, we used simple statistics to report the number of students in each paper and the number of the times the courses have been offered.

Last but not the least, for RQ 9.1 and RQ 9.2, we use a more sophisticated approach for the analyses, i.e., qualitative coding, which is a systematic qualitative data analysis approach [83], to synthesize challenges in and insights for testing education, as reported in the papers. Thus, for these two RQs (RQ 9.1 and RQ 9.2), the evidence synthesis approach that we selected to use is systematic literature review. Our qualitative coding approach is explained with some examples in Section 4.5.

**Table 5: Systematic map developed and used in our study**

| RQ | Sub-RQ | Attribute | Categories/metrics | Single or Multiple selections | Answering approach |
|---|---|---|---|---|---|
| 1 | 1.1 | Contribution type | {Pedagogical Approaches (how to teach better), Course-ware / new course proposal, (Proposing) a specific tool for testing education, Gamification for testing education, Status / overview / trends, Empirical study only, "Other" contributions} | Multiple | Systematic mapping |
| | 1.2 | Clusters of papers on same topic, often by the same authors team | List of authors, to be later used to find paper clusters | Multiple | Clustering |
| 2 | - | Research-method type | {1-Proposal with very little experience, 2-Experience / Informal evaluation, 3-Experiment / empirical study, Review (of practices) } | Single | Systematic mapping |
| 3 | - | Data source (if any) used for the evaluation | {Survey (e.g., from students), Interview with students, Student-written tests, Other} | Multiple | Systematic mapping |
| 4 | - | Research questions or hypothesis, studied in the papers | Research questions or hypothesis, raised and studied in each paper | Multiple | Merging all the data |
| 5 | - | Type of test activities covered in the course | {Generic software testing, Test process, Test-case Design (Criteria-based), Test-case Design (Human knowledge-based), Test Automation, Test Execution, Test evaluation, Other test activities} | Multiple | Systematic mapping |
| 6 | - | Context (classes) under study | • The number of students taking the course<br>• The number of the times the course has been offered | Single | Simple statistics |
| 7 | 7.1 | Independent testing course or integration of testing across one or more other courses: Share of papers in each approach | {Independent testing courses, Teaching testing in one programming /SE course, Teaching testing across SEVERAL programming courses, Not clear or N/A} | Single | Systematic mapping |
| | 7.2 | Goal (purpose) of the evaluations if testing is spread across courses/program | Goal of each paper w.r.t. the issue | Multiple | Merging all the data |
| 8 | - | Theories and theory use in software-testing education | The list of theories (if any) used in a given paper and for the purpose(s) they are used for | Multiple | Their raw text |
| 9 | 9.1 | Empirical evidence / findings- Challenges in testing education | Explanations about challenges in testing education | Multiple | Qualitative coding (systematic review) |
| | 9.2 | Empirical evidence / findings- Insights for testing education | Explanations about insights when teaching testing | Multiple | Qualitative coding (systematic review) |

### 3.6 DATA EXTRACTION PROCESS AND DATA SYNTHESIS

Once the classification scheme was developed, we first conducted a "pilot" data extraction phase in which each researcher extracted data from five papers and we then peer-reviewed the extracted data to cross validate and refine our data extraction approach. This was done to ensure homogeneity of the work by different team members and also to ensure the quality of our analyses. Having completed a pilot phase, we then partitioned the pool of papers among all the researchers.

Each researcher extracted and analyzed data from the subset of the papers assigned to the researcher. When recording the extracted data in a master spreadsheet, we included additional comments in each paper to make explicit why the given paper was classified for each attribute in the specific way (see the example in Figure 6).

Figure 6 shows a snapshot of our online spreadsheet that we used to enable collaborative work and classification of papers with traceability comments. In this snapshot, classification of papers for RQ 2 (research method) is shown. One researcher has placed the exact phrase from the [P31] as the comment to facilitate peer reviewing and also quality assurance of data extractions, afterward during peer-reviewing phase.

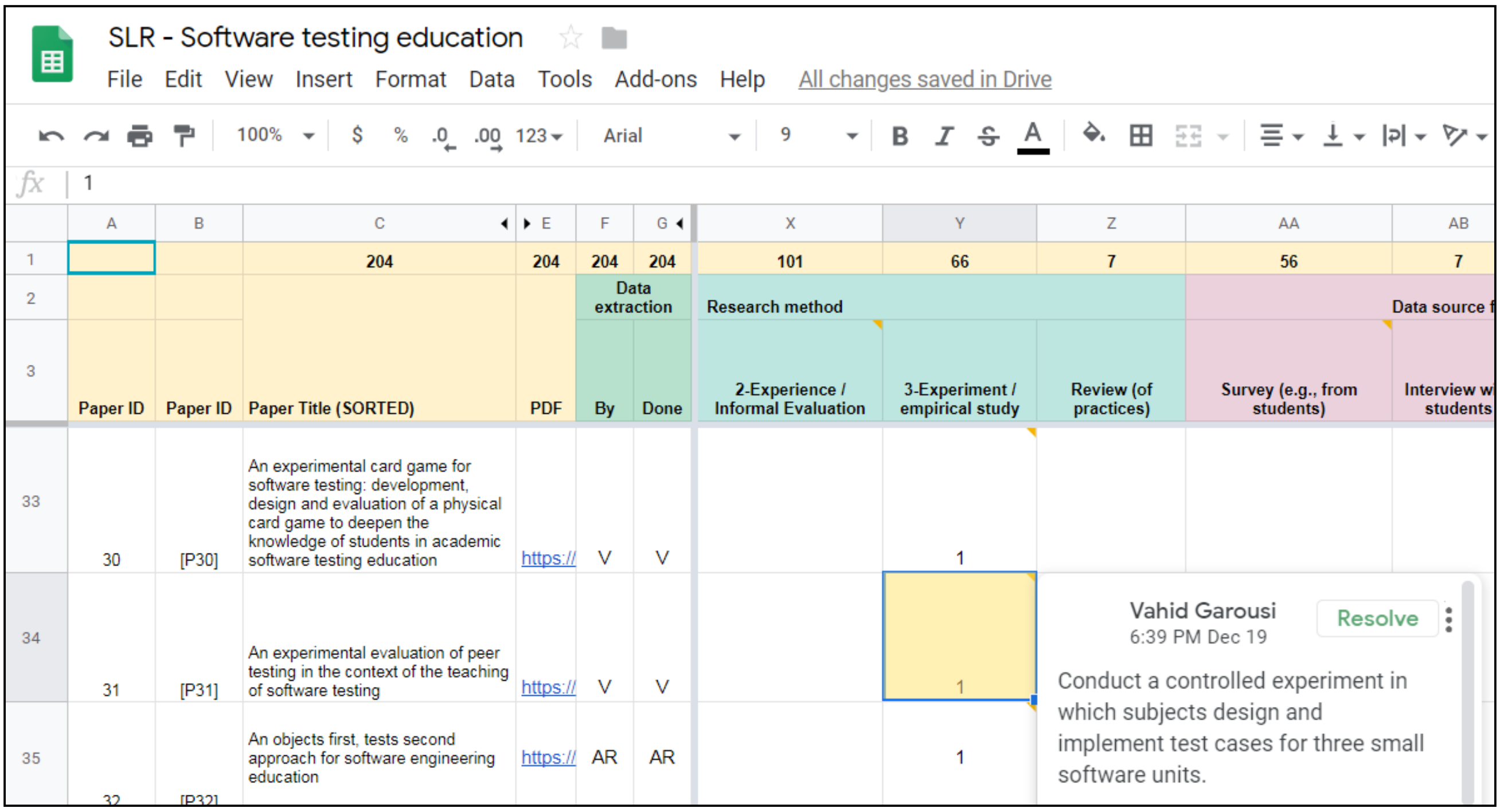

**Figure 6- A screenshot from the online repository of papers (goo.gl/MhtbLD).**

After all researchers finished their data extraction, we conducted systematic peer reviewing in which researchers peer reviewed the results of each other's analyses and extractions. In the case of disagreements, discussions were conducted. This was done to ensure quality and validity of our results.

When synthesizing and reporting challenges for RQ 9.1, there were cases that we did not necessarily agree that the reported "challenge" is actually a challenge. For example, [P116] reported that: "*Many of the students found JUnit to be too complicated for them*", which is quite subjective to interpret as an actual challenge. We therefore synthesized and report the challenges as they have been reported in the papers. In reporting these challenges, we do not necessarily advocate for them as challenges.

As shown in Table 5, to address two of the study's RQs, RQ 9.1 and 9.2, we conducted qualitative coding of data. To choose our method of synthesis, we carefully reviewed the research synthesis guidelines in SE, e.g., [84-86], and also other SLRs which had conducted synthesis of results, e.g., [87, 88]. According to [84], the key objective of research synthesis is to evaluate the included studies for heterogeneity and select appropriate methods for integrating or providing interpretive explanations about them [89]. Since the primary studies have not reported similar-enough measures with respect to interventions and quantitative outcome variables, meta-analysis was not applicable nor possible. As we examined our papers, we concluded that the best applicable method for RQ 9.1 and 9.2 was qualitative coding using "open" and "axial coding" [83].

In our initial screening of the extracted data for "challenges" and "insights", a few initial groups emerged as a major challenge, e.g., testing often not well accepted among students. During the rest of our qualitative data analysis process, we found out that the initial list of challenges and insights had to be expanded, thus, the rest of the factors emerged from the papers. The creation of the new factors in the "coding" phase was an iterative and interactive process, which was conducted by one researcher and peer reviewed by all others. Basically, we first collected all the factors related to questions RQs 9.1 and 9.2 from the papers. Then we aimed at finding factors that would accurately represent all the extracted items but at the same time not be too detailed so that it would still provide a useful overview, i.e., we chose the most suitable level of "abstraction" as recommended by qualitative data analysis guidelines [83].

Figure 7 shows a screenshot from the datasheet, in which an example of qualitative coding to answer RQ 9.1 (challenges when teaching testing) is shown. From the example paper in this case, [P11], the reported challenges were extracted and grouped into two under column "Raw phrases": (1) "*It is challenging to teach software testing in a way that is engaging for students, and to ensure that they practice effective testing sufficiently*"; "*maintaining student interest*"; (2) "*setting the appropriate complexity for students*". We then coded the raw phrases in the right-hand side categories and removed them from the raw phrases list after being coded (i.e., by "consuming" them). In the case of this example (Figure 7), the raw data were coded under two categories as visualized.

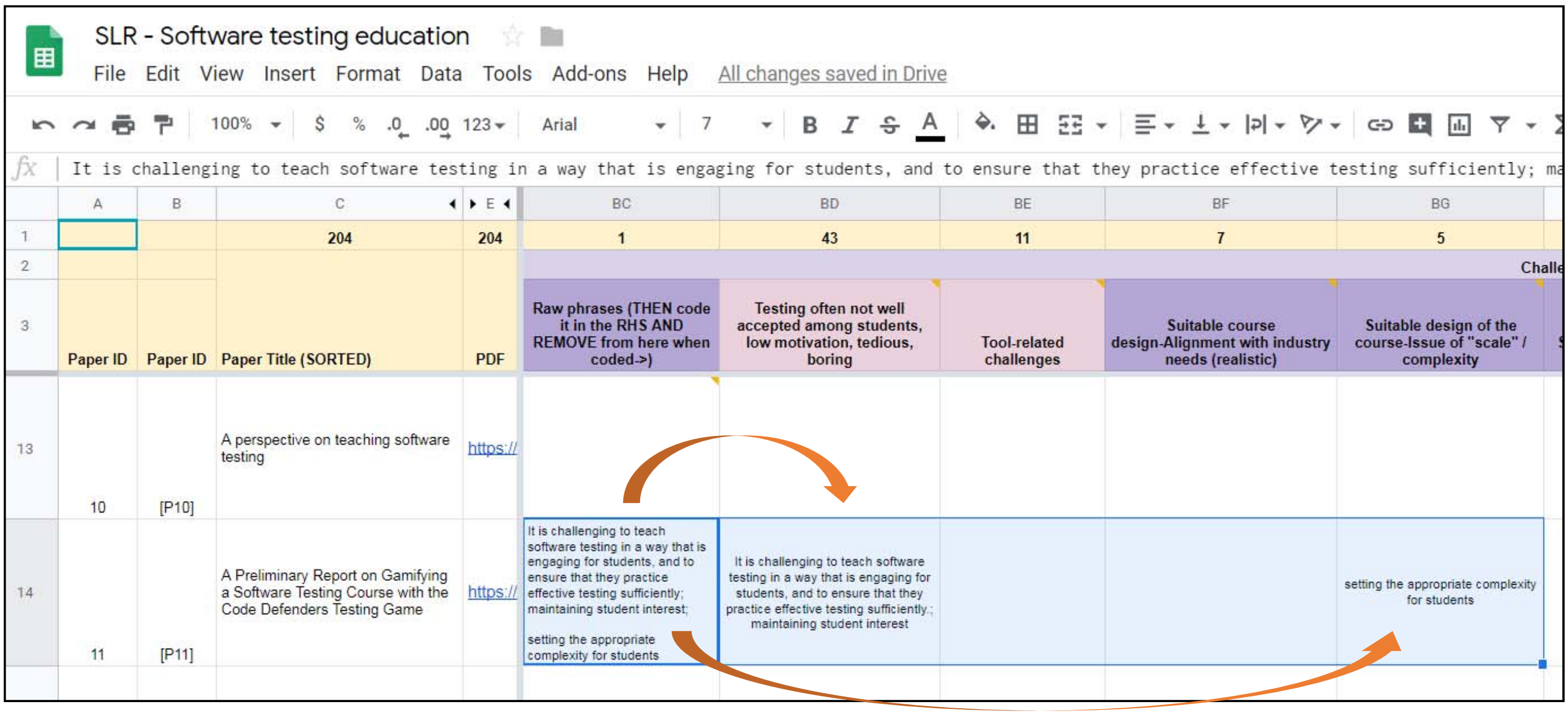

**Figure 7: A screenshot from the datasheet showing qualitative coding of data to answer RQ 9.1 (challenges when teaching testing)**

## 4 Findings of the SLM

Through Section 4, we present results of the study's RQs.

### 4.1 RQ 1-Classification of studies by contribution types

RQ 1 consists of RQ 1.1 and RQ 1.2, as presented next.

#### 4.1.1 RQ 1.1-Contribution types and their frequencies

Figure 8 shows the classification of studies by contribution types. As we discussed in the structure of the systematic map (Table 5), since each paper could have multiple contribution types, it could thus be classified under more than one category in Figure 8. For example, [P23] presented an automated tool to help students learn how to write better test cases. The system was implemented for Python programs, but as the paper explained, "*the pedagogical principles underlying it transcend any particular language*". Thus, the paper was marked under both Pedagogical approaches and Proposing a specific tool for teaching testing.

As we can see in Figure 8, the top three types of contributions are: (1) Pedagogical approaches (how to teach better) with 100 papers (49.0% of the pool), (2) Proposing a specific tool for testing education, with 62 papers (30.4% of the pool), and (3) Course-ware / new course proposal (except tool) with 49 papers (24.0% of the pool). We discuss below a summary of each category by referring to a few example papers in that category.

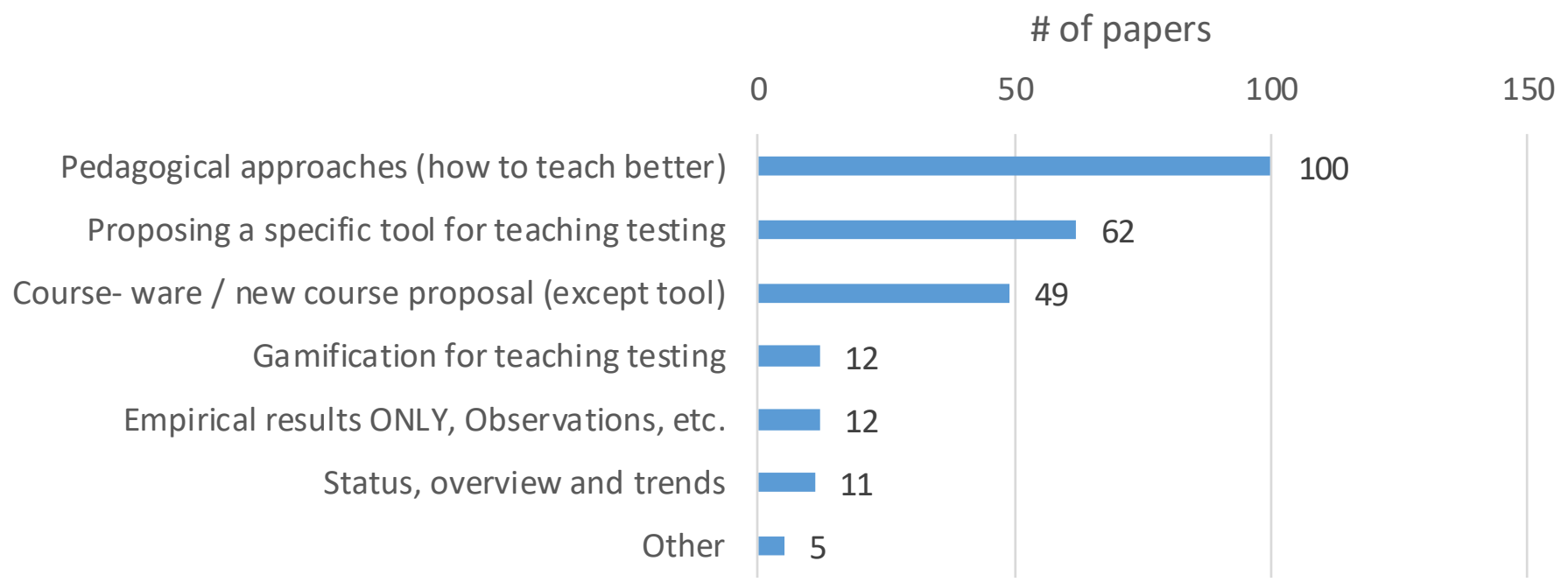

**Figure 8- Classification of studies by contribution types**

Multiple studies within the **Pedagogical approaches** category discuss when to introduce software testing. We find papers reporting benefits of introducing testing early in introductory programming courses, e.g. [P177]: *"(…) the emphasis on testing has been qualitatively beneficial for students in the introductory Java courses"*. We also find papers describing challenges in introducing testing early: *"Testing methods are impossible to understand by students without programming experience, and their depth of programming practice is also a factor"* [P75].

It may be hard for a student to see the value of software testing when a coding project is of limited size. Multiple studies in our pool therefore recommend using free or open-source projects (F/OSS) when we teach software testing, e.g. [P195] (on using a real-world project in a software testing course): *"We have witnessed a significant increase in student enthusiasm in software testing as a subject and a discipline"*. F/OSS projects will also give students practice in testing code they have not written themselves. This may also be achieved by having students write tests for code produced by fellow students [P12, P17, P23, P73, P95, P161, P199]. It might be more motivating to find other people's bugs, rather than your own: *"When testing their own code, students are less motivated to find bugs, as bugs expose their own failure to develop a correct program"* [P23].

So within the pedagogical approaches category we find multiple suggestions for increasing the motivation for software testing - motivation other than simply having testing as an evaluation criterion when students produce software code. *"Students need to directly experience benefits from writing test suites. Requiring students to write test cases simply because test suite quality will be graded does not help students learn the value of testing"* [P85].

As software testing may be a difficult topic for the inexperienced programmer, the educator may use **Specific tools** for support in the learning environment. As an example, when TDD is used in an introductory course, *WebIDE* can guide the students through a pre-defined path of steps in order to force them to write tests and specifications before implementing solutions [P145]. Many tools have been developed for software testing in education over several years. As some of these tools have multiple papers associated with them, we describe them further in the following section.

The **Course-ware/new course proposal** category holds papers with descriptions of courses or modules where software testing is involved. The papers may include course evaluations and lessons learned from delivered courses, or they may present new ones. These papers can provide valuable information to educators who want to include software testing practices to existing courses or when building a new course. Some papers also include URLs where course materials can be accessed. As an example, [P33] presents software testing laboratory courseware with descriptions of all labs in a 13-week course on software testing. Within the descriptions, we find learning objectives, different test activities, tools and languages involved along with student evaluations. The entire repository for the course is available online through a provided URL.

**Gamification** may increase motivation for students studying software testing. Educators can introduce the typical gamification elements such as reward points, badges, and leader boards in a learning environment [P82]. Gamification can also be achieved through educational games: *HALO* (Highly Addictive sociaLly Optimized Software Engineering) [P5, P133], *Code Defenders* [P11, P48, P83, P157, P163], *PlayScrum* (a physical card game) [P30], *Bug Hide-and-Seek* [P42], *Testing Game* [P64] and *U-TEST* [P66].

Studies within the **Empirical results** category report from investigations performed in a software testing education environment. Some papers report on achieved improvements after course adjustments. The improvement can be in code quality after introducing unit testing in a programming course [P34] or in students' conceptual understanding of software testing after introducing Software Testing Computer Assistant Education (STCAE) [P49]. Within the category, we also find investigations into student software testing behavior: The quality of student-written tests in terms of the number of authentic, human-written defects those tests can detect [P63], or software testing behaviors that students exhibited in introductory computer science courses [P65]. Code coverage alone will not determine if student code is tested well enough. We may use mutation analysis to fix some problems of code coverage in student assessments [P115].

In the **Status, overview and trends** category, we find papers investigating the current state of software testing in education in different parts of the world: Australia [P14], Canada and America [P54], Hong Kong [P136], South Africa [179] and Brazil (and abroad) [152]. We also find papers describing graduates' knowledge gaps in software testing according to industry needs [P15, P179, P181]. A common conception within the papers in this category is that "*More testing should be taught*" [114]. As described in [15]: "*In general, results indicated a deficiency for all testing topics in practice activities. In particular, there were also negative gaps in topics such as test of web applications, functionality testing and test case generation from client requirements/user stories*".

#### 4.1.2 RQ 1.2: Clusters of papers: on similar topics and by the same team of authors

When extracting data from our pool of papers, we observed that some papers were similar regarding both topic and authors involved. When we found two similar papers, we kept both only if each of them provided new insight. Some papers were easily recognized as linked together as they all described a common tool or artefact. We found two tools and one game in a substantial amount of papers: *WReSTT-Cyle* (7), *Web-CAT* (8) and *Code Defenders* (5).

Papers on **WReSTT-Cyle** (Web-Based Repository of Software Testing Tutorials: A Cyberlearning Environment) appear from the year 2010. "*The main objective of the online portal is to increase the number of users at academic institutions that currently have access to vetted learning materials, including tutorials on software testing tools, that support the integration of testing into programming courses*" [P196]. Subsequent papers describe insight into using the repository in software-testing education [P8, P82, P89, P101, P103, P203]. "*(…) WReSTT has evolved into a collaborative learning environment with social networking features such as the ability to award virtual points for student social interaction about testing*" [P8]. The learning environment is now called **SEP-CyLE** (Software Engineering and Programming Cyberlearning Environment).

Within our pool of papers, we find eight involving **Web-CAT** from the year 2003 to 2014 [P3, P65, P87, P90, P96, P132, P198, P200]. "*Web-CAT is a web application with a plug-in architecture that can provide a variety of services for students. Its Grader plug-in provides a highly configurable and customizable automated grading and assessment service*" [P87]. A motivation behind *Web-CAT* is to "*(…) encourage students to adhere to Test-Driven Development (TDD)*" [P3]. It is an adaptive feedback system where students can submit code (multiple times) and receive automated feedback. "*With each submission, students receive feedback including analysis of code style, correctness, and testing quality. The correctness is calculated by the percent of instructor-written tests (obscured from the student) that pass when run against the students' code. Testing quality is represented by code coverage, or the percent of statements, conditionals, and branches executed by the student's own unit tests. Additionally, Web-CAT highlights the student's code to reveal where coverage is lacking. After receiving feedback, students may correct their code and resubmit to Web-CAT without punishment*" [P3]. The student submissions in *Web-CAT* provide insight into common student testing behaviour, and how the behaviour may change when feedback is provided by *Web-CAT*.

The most recent cluster of papers (2016-2019) involves a game for mutation testing: **Code Defenders** [P11, P48, P83, P157, P163]. The papers describe design and implementation details for the tool [P48] and how we can use it for educational purposes [P157] for different software testing methods [163]. Subsequent papers provide empirical data from the game in actual use and how it can be integrated in a course on software testing [P83]. The game includes two game modules: "*(…) duel, where each game consists of one attacker and one defender competing in a turn-based fashion, very much like a traditional board game; and battle, where each game consists of a team of defenders and a team of attackers that play without turns. Attackers use a code editor to introduce artificial faults. This resembles the idea of mutation testing, where artificial defects are produced automatically*" [P11]. The initial reports are promising with students enjoying the game while improving their software testing skills.

### 4.2 RQ 2-CLASSIFICATION OF STUDIES BY RESEARCH METHOD TYPES

In SLMs, e.g., [51-55], it is also common to classify primary studies by their types of research methods. As the structure of the systematic map (Table 5) showed, based on established review guidelines [14, 48-50], we classified the papers as follows: (1) Proposals of ideas or approaches for testing education, with no explicitly-mentioned experience in the paper, (2) Experience / informal evaluation, and (3) Experiment / empirical study.

Figure 9 shows the breakdown of the primary studies by the above research facets. As we can see, more than half of the papers (116 of 204) are experience papers. It is particularly encouraging to observe there are many empirical studies in the pool.

For the 10 papers which were in the form of proposals only, we observed that although the authors had not shared any experience of applying the ideas / approaches in their testing courses, it was clear that authors were active testing educators, and thus, at least implicitly, they had tried or were going to try the ideas / approaches in their testing courses. We designed the categorization of three research method types shown in Figure 9 to be able to interpret them as three "levels" of evidence: #1, #2, and #3. Studies which reported empirical studies can generally be seen as having the highest (most rigorous) level of evidence.

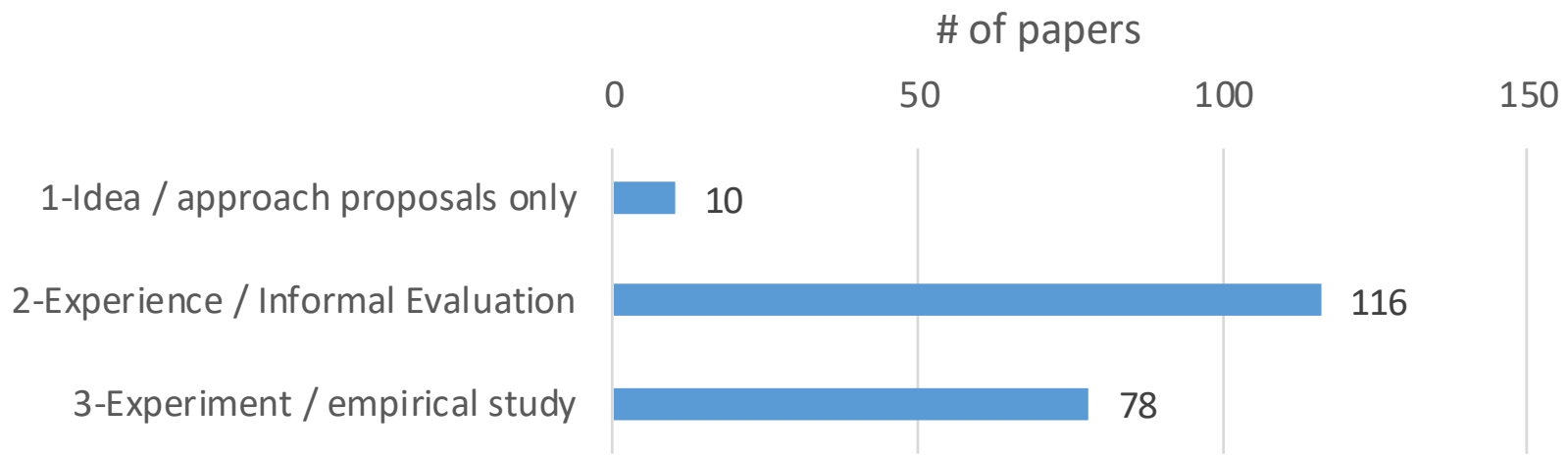

**Figure 9- Breakdown of primary studies by research-method types**

With the above breakdown, we can relate the research-methods used in the studies to a hierarchy of evidence for software engineering research, as proposed by [90], which is shown in Table 6. (We could have used the following more detailed hierarchy of evidence in our work, but the hierarchy came to our attention after we had carried out our work.)

**Table 6: A hierarchy of evidence for software engineering research, as proposed by [90]**

| Type of study | | Level | Evidence | Categorization of research methods in this paper |
|---|---|---|---|---|
| Secondary or filtered studies | | 0 | Systematic reviews with recommendations for practice; meta-analyses | - |
| Primary studies | Systematic evidence | 1 | Formal or analytic results with rigorous derivation and proof | - |
| | | 2 | Quantitative empirical studies with careful experimental design and good statistical control | - |
| | Observational evidence | 3 | Observational results supported by sound qualitative methods, including well-designed case studies | (3) Experiment / empirical study |
| | | 4 | Surveys with good sampling and good design; field studies; data mining | |
| | | 5 | Experience from multiple projects, with analysis and cross-project comparison; a tool, a prototype, a notation, a dataset, or another artifact (that has been certified as usable by others) | (2) Experience / informal evaluation |
| | | 6 | Experience from a single project: an objective review of a specific project; lessons learned; a solution to a specific problem, tested and validated in the context of that problem; an in-depth experience report; a notation, a dataset, or an unvalidated artifact | |
| No design | | 7 | Anecdotes on practice; a rule of thumb; an evaluation with small or toy examples; a novel idea backed by strong argumentation; a position paper | (1) Proposals of ideas or approaches for testing education, with no explicitly-mentioned experience in the paper |

## 4.3 RQ 3- Data sources for evaluations

Different papers collected and used different data sources for evaluations. We classify and show the frequencies of those data sources in Figure 10. The majority of papers collected data from students via surveys and then used those data to evaluate the testing approaches presented in the papers. This is expected, as surveys are easier to perform and to analyze, compared to the other sources. Furthermore, surveys can be included in the regular evaluation forms of the courses, usually done at the end of the course to provide feedback. This is a common practice in many universities.

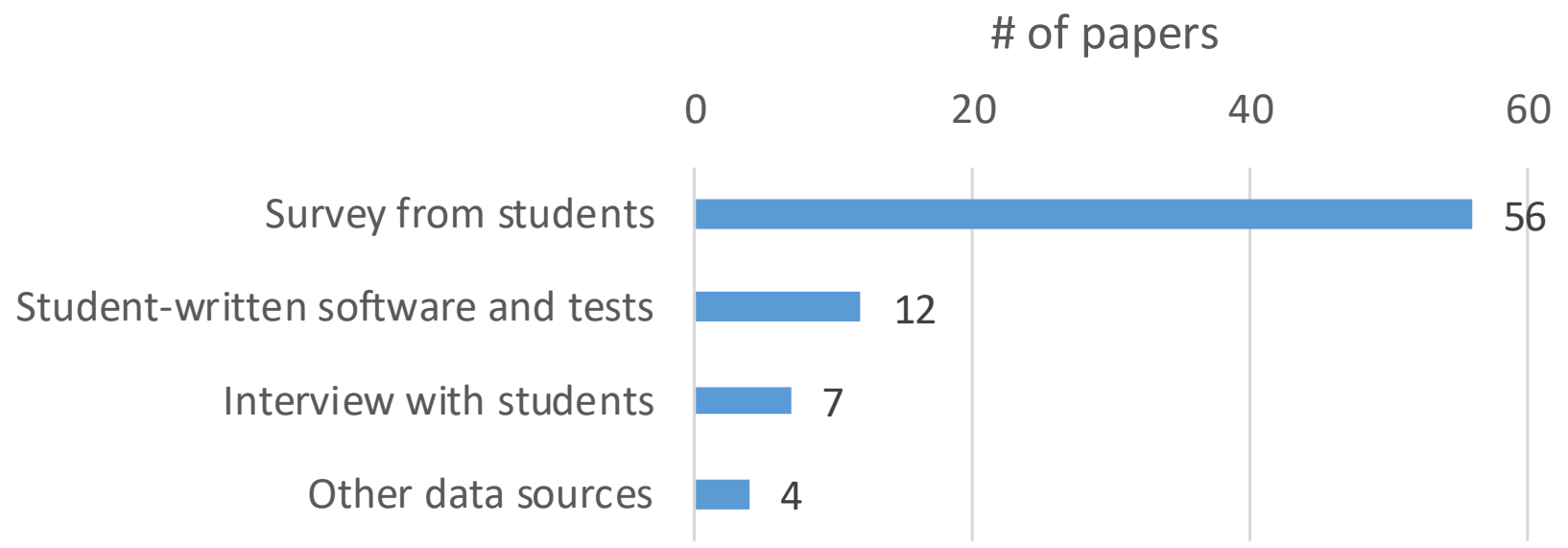

**Figure 10- Data sources for collecting evaluation data**

In 12 papers, for courses with practical exercises and exams, evaluations and conclusions were done based on production-code and automated tests developed by the students. For example, "*To assess how ProgTest was able to help students, we record the status of all students' submissions*" [P158]. The quality of the implemented software was also used as a metric to evaluate the impact of testing education, e.g., "*[…] while producing code with 45% fewer defects per thousand lines of code*" [P201].

To gain an even better understanding of how the teaching of testing impacts the students, interviews were reported in some papers. For example, in [P81]: "*a series of guided group interviews with project teams was conducted. It included 113 students in 39 teams*". But as interviews are often time-consuming to perform, compared to surveys, it was not a common method among the analyzed papers. There were also other less common data sources, such as online quizzes [P25].

### 4.4 RQ 4- Research questions or hypothesis, studied in the papers

In this section, we analyze the RQs and hypotheses studied in the papers we reviewed. Quantifying the number of papers that explicitly investigate RQs and hypotheses provides a complement to Section 5.2 and Figure 9, and provides another indication of the amount of *research* being conducted on software-testing education. Knowing about the RQs investigated in those papers also helps others to explore similar or contrasting research directions in their research in the future.

46 out of the 204 papers (%22.5) explicitly stated RQs or hypotheses, suggesting that about a fifth of the papers we reviewed constitute research into software-testing education. From those 46 papers, we identified 121 RQs and hypotheses (H). Those RQs and hypotheses are summarized in Table 7. The vast majority of the 46 papers investigate RQs rather than hypotheses. The RQs and Hs are numbered sequentially, although we use H0 to denote null hypotheses. To clarify some of the items in Table 6, we provide additional contextual information in parentheses where needed.

We can see in Table 7 that a variety of RQs have been evaluated in previous studies. Using Easterbrook et al.'s [91] classification, we see that there are different types of RQs in the list, e.g., exploratory RQs, relationship RQs and causality RQs. The number of RQs suggests that a considerable amount of empirical data have been gathered and reported in previous research. While we take an initial step in this paper in synthesizing some of that evidence (see Section 5.8), further work is needed to comprehensively synthesize evidence from similar or related RQs in Table 7.

**Table 7: Research questions or hypotheses investigated by different papers**

| Paper ID | RQs |
|---|---|
| [P8] | • RQ1: Does the integration of SEP-CyLE have a significant and quantitative impact on the programming and testing knowledge gained by the students?<br>• RQ2: Are SEP-CyLE testing-related assignments aligned with the needs and interests of the learners in understanding underlying programming concepts? |
| [P9] | • RQ1: Is it possible to take existing materials and tack on a TDD approach?<br>• RQ2: Is giving credit to tests the best way to teach TDD? Do students write more, higher quality tests if they get feedback through grades on tests?<br>• RQ3: Does the TDD approach affect the amount of time spent on projects, since students have to write test-code?<br>• RQ4: Does writing tests lead to higher quality code with respect to the number of acceptance tests passed?<br>• RQ5: If in-class examples are developed using a TDD approach, does it have a higher impact on students than those who do not see testing in class? |
| [P13] | • RQ1: The main question that needs to be answered is whether (1) investing in an expensive (25.000 euros) physical infrastructure really creates a substantial positive effect on ST learning.<br>• RQ2: A less critical question is whether one month offers enough time to overcome the non-technical background of CS students and really get focus on testing. |
| [P15] | • RQ1: What are the knowledge gaps in testing topics faced by graduates with respect to industry needs? |
| [P18] | • RQ1: Can unit testing improve the quality of human computer interaction projects?<br>• RQ2: When introducing unit testing, what additional steps must be taken to ensure a positive learning experience?<br>• RQ3: What potential for regression of students' unit testing model is possible, and how can that potential be mitigated? |
| [P20] | • Alternative hypothesis: post-test scores are significantly higher than pre-test scores on average. (Pre-test and post-test covered all the learning objectives of the course described in the study. The description includes the pedagogical approach taken.) |
| [P26] | • RQ1: Do unit-testing practices in CS1 assignments and labs really improve code quality?<br>• RQ2: Do CS1 students enjoy writing test cases?<br>• RQ3. Do unit-testing practices in CS1 enhance the student's learning process? |
| [P27] | • RQ1: Students' attitude toward accepting non-traditional educational module is more positive than toward accepting traditional one?<br>• RQ2: If subjects were given training using non-traditionally-produced educational module would behave more uniformly, in the sense of fault detection rate, than if they were given training using traditionally-produced module?<br>• H0: There is no difference in the fault detection rates uniformity of subjects given training using non-traditionally-produced module as compared to subjects given training using traditionally-produced module. |
| [P28] | • H1: Students who received test sets T1 and T2 will produce higher quality programs than students who received only the program specifications. |
| [P31] | • RQ1: Is peer testing more effective than individual testing for the construction of test cases?<br>• RQ2: Is peer testing more efficient than individual testing for the construction of test cases? |
| [P38] | • RQ1: How can (i) participation and (ii) performance in agile testing be measured? |
| [P40] | • RQ1: Which test quality measures actually assess how much of the expected behavior is checked by the tests?<br>• RQ2: What are the practical obstacles of using identified test quality measures in an educational setting?<br>• RQ3: How can we resolve the obstacles to apply the measures in classroom tools?<br>• RQ4: Which approach is more appropriate for open-ended assignments?<br>• RQ5: What measure works better for close-ended assignments?<br>• RQ6: What combination of the approaches works well as a hybrid measure to separately evaluate tests of the assignments having variable amounts of design freedom? |
| [P41] | • RQ1: How many bugs does each team find? (In a software testing competition using Bug Catcher – a web-based system for running software testing competitions)<br>• RQ2: Do the students recommend this event for future students? (The software testing event using Bug Catcher)<br>• RQ3: Do the students report an increased interest in Computer Science? (After the event)<br>• RQ4: What are suggestions for improving the system? (Bug Catcher) |
| [P46] | • Hypothesis: Including software security testing techniques as part of the typical software testing exercises used in CS classrooms will expand students' programming toolset and make them better equipped to tackle programming tasks.<br>• RQ1: Were student submissions unique? (Students wrote both submissions (defence programs) and test cases (attack programs) for an assignment given in an introductory security class.)<br>• RQ2: Do multiple attacks benefit performance? (Did students acquire a better score if they submitted multiple attack submissions?)<br>• RQ3: What accounts for the difference between max and overall SAQ? (SAQ score: student attack quality as a student's overall ability to attack all monitors.)<br>• RQ4: Are attack/defense abilities correlated? |
| [P47] | • RQ1: Whether either checked coverage or object branch coverage is a better indicator of test suite quality than a number of alternative measures—that is, is either a more accurate predictor of a test suite's ability to detect faults? |
| [P51] | • RQ1: How to make writing tests more reasonable in the educational context? |
| [P63] | • RQ1: How good are student-written tests at finding real bugs?<br>• RQ2: How much variation is there in the software tests written by students? |
| [P64] | • RQ1: Does the Testing Game have good quality regarding motivation, user experience and learning, from students' point of view? (Testing Game: an educational game addressing the following topics: functional testing, structural testing and mutation testing.)<br>• RQ2: Does the Testing Game have good usability from the student's point of view? |
| [P66] | • RQ1: Is there any different in relative learning in the game higher than in the group that did not play?<br>• RQ2: Is the education game considered appropriate in terms of content relevancy, correctness, and degree of difficulty? Is the game considered engaging? |
| [P67] | • RQ1 (Testing Strategies): How did students test their software products?<br>• RQ2 (Enabling and Inhibiting Factors): What factors supported students in testing methodically and what factors hindered them?<br>• RQ3 (Testing Attitude): What did students think of testing methodically?<br>• RQ4 (Testing in the SWP process): How did students incorporate testing in their engineering process? |
| [P71] | • RQ1: Can the mutation testing criterion facilitate the learning process of novice students in programming courses?<br>• RQ2: What are the trade-offs and recommendations of using mutation testing to support the learning process in programming courses? |
| [P72] | • RQ1: Can ST knowledge help developers improve their programming skills in terms of delivering more reliable implementations?<br>• RQ2: Does ST knowledge impact on the effort invested by developers on their implementations?<br>• RQ3: Does ST knowledge impact on the complexity of the produced code? |
| [P74] | • RQ1: What are the beneficial on-line services for successful testing course?<br>• RQ2: To what extent can a technically challenging CSE course be offered online? |
| [P79] | • RQ1. Is there a significant difference between the students' performances under different testing techniques?<br>• RQ2. Does there exist a noticeable relationship between the tests results under different testing techniques?<br>• RQ3. What is the importance of the programming background when applying different testing techniques?<br>• RQ4. What is the influence of the gender factor on success in software testing assessments?<br>• RQ5. How do various teaching strategies over the years affect the exam results in the software testing? |
| [P80] | • H1: Students will rate importance of skills and their corresponding strengths with a positive correlation<br>• H2: Students will rate helpfulness of and their adherence to behaviors with a positive correlation<br>• H3: Students more likely to adhere to TDD principles will rate TDD's helpfulness more positively<br>• H4: Students with higher programming anxiety (according to WTAS) will adhere less to starting work early and to principles of TDD<br>• H5: Students with higher programming anxiety will rate Web-CAT as more helpful<br>• H6: Students with higher evaluation anxiety (according to BFNES) will rate Web-CAT as less helpful. |
| [P83] | • RQ1: How do students engage with the game? (The Code Defenders game: Students compete over code under test by either introducing faults ("attacking") or by writing tests ("defending") to reveal these faults.)<br>• RQ2: Does student performance improve over time?<br>• RQ3: Does student engagement correlate with exam grades?<br>• RQ4: Do students appreciate using Code Defenders in class? |
| [P84] | • RQ1: Which testing tools and technologies are most used in the industry?<br>• RQ2: What are the current issues related to testing in the industry?<br>• RQ3: How should the learning goals, teaching methods and evaluation methods in a software testing course constructively aligned with current industry practices? |
| [P88] | • RQ1: Is the level of CS program exposure related to the quality of test cases generated with black-box and white-box methods by undergraduate and graduate students? |
| [P89] | • RQ1: If the availability and knowledge of the use of code coverage tools positively impacts and increases students' propensity to improve the quality of their black-box test suites.<br>• RQ2: If an increase in code coverage during white-box testing results in an increase in the number of bugs students find during testing.<br>• RQ3: If students find WReSTT a useful learning resource for testing techniques and tools.<br>• RQ4: If students find that WReSTT supports collaborative learning. |
| [P90] | • H1: The experimental group will have significantly greater average TMSM and average coverage than the control group. (TMSM: average test-methods-per-solution-method. The experimental group used a plugin for Web-CAT that provides adaptive feedback based on how well the student is adhering to incremental unit testing.)<br>• H2: The experimental group will have significantly greater project correctness and coverage scores than the control group.<br>• H3: The experimental group's average TMSM and average coverage will increase over time relative to the control group's average TMSM and average coverage trends.<br>• H4: Students' perceptions of the helpfulness of test-first and unit testing will have a positive correlation with their self-reported adherence to the same behaviors.<br>• H5: The experimental group will value the helpfulness of test- first and unit testing behaviors significantly higher than the control group.<br>• H6: The experimental group will score significantly lower on WTAS (project anxiety) scale relative to their BFNES (fear of negative evaluation) scale when compared to the control group.<br>• H7: The experimental group will respond more positively to following TDD in the future than the control group. |
| [P93] | • RQ1: Can TDD be integrated into early programming courses with minimal effort on the part of instructors?<br>• RQ2: What effect does the grading of test-code have on students' tests?<br>• RQ3: What effects does TDD have on quality of code and productivity of students? |
| [P95] | • H1: Written test cases based on pair programming increase the number of killed mutants.<br>• H2: Written test cases based on pair programming provide better code coverage. |
| [P115] | • RQ1: What are the possible strengths and weaknesses of mutation analysis when compared to code coverage based metrics?<br>• RQ2: Can mutation analysis be used to give meaningful grading on student-provided test suites requested in programming assignments? |
| [P123] | • RQ1: Can POPT help students to obtain more correct implementations than traditional approach based on blind<br>• testing? (POPT: A Problem-Oriented Programming and Testing Approach for Novice Students)<br>• RQ2: Do students adopting POPT submit fewer versions than the ones using traditional approach?<br>• RQ3: Do POPT programmers spend more time to deliver the implementation than traditional programmers? |
| [P125] | • RQ1: What common mistakes do students make when learning software testing? |
| | • RQ2: Which software testing topics do students find hardest to learn?<br>• RQ3: Which teaching methods do students find most helpful? |
| [P126] | • RQ1: Can we lead students towards the habit of writing tests in software projects using introductory programming exercises?<br>• RQ2: Can these exercises be implemented in a highly automated, yet student-centered manner? |
| [P137] | • RQ1: Is the effectiveness in the detection of defects affected by the use of a CVE? (CVE: collaborative virtual environment) |
| [P143] | • RQ1: Does the use of code coverage tools motivate students to improve their test suites during testing?<br>• RQ2: Do the results generated by the code coverage tools support the subsumes relation between branch coverage and statement coverage, i.e., does branch coverage subsume statement coverage?<br>• RQ3: Do students find WReSTT a useful learning resource for testing techniques and tools?<br>• RQ4: Do students find the features in WReSTT support collaborative learning? |
| [P145] | • H1: Students who used WebIDE perform better on programming tasks than students who used traditional static labs.<br>• H2: Students who used Web-IDE spend more time on labs (because of the lock-step aspect) than students who used traditional static labs. |
| [P157] | • H1: Through the use of a mutation testing game, students will be able to grasp all relevant mutation testing concepts while having fun, and in the end become better software developers and testers, who produce higher quality software. |
| [P158] | • H1: The proposed tool (ProgTest) helps novice programmers to increase the quality of their programs and test suites. |
| [P167] | • RQ1: Does the use of Coding Dojo methodology to teach TDD improve the code coverage of students when compared to solo programming?<br>• RQ2: Does the use of Coding Dojo methodology improve motivation and grow the interest in learning TDD when compared with solo programming? |
| [P168] | • RQ1: Does TFD have any effect on the learning process?<br>• RQ2: Does it impact the way inexperienced students code?<br>• RQ3: Will this experience have long-lasting effects on students? |
| [P182] | • RQ1: Can ST knowledge help developers improve their programming skills in terms of delivering more reliable implementations?<br>• RQ2: Does ST knowledge impact on the effort invested by developers on their implementations?<br>• RQ3: Does ST knowledge impact on the complexity of the produced code? |
| [P190] | • H0: Is the code correctness using instructor-provided test cases equal to that with student-written test cases? |
| [P204] | • H0: There would be no significant difference between the performances in terms of scores of students on the first programming assignment from 2001 and 2003. (To assess the effectiveness of Web-CAT and TDD.) |

### 4.5 RQ 5- TECHNICAL ASPECTS OF TESTING: TYPE OF TEST ACTIVITIES COVERED IN THE COURSE(S)

For RQ 5, we wanted to assess the types of test activities covered in the educational courses. We acknowledge that the papers in our set of primary studies are a (very) small sub-set of all testing courses taught at universities world-wide. Clearly, not all testing educators publish education papers about their teaching activities and experiences. Thus, RQ5 is not aiming to provide a world-wide view on the type of test activities covered in "all" testing courses world-wide, but rather only in the set of the primary studies under review in this work. In other words, we recognize that there is an issue of generalizability for our analyses of RQ5.

To classify test activities, we re-used the process model for testing from a previous paper, as shown in Figure 12 (from [92]). Note that a paper could be classified according to more than one type. Figure 11 summarizes the frequencies of type of test activities covered in the courses. Four types of test activity are each studied by approximately 40% of the 204 papers we examined, i.e., generic software testing, test-case design, test automation and test execution. Relatively speaking, test evaluation has not been investigated (41 instances in 204 papers). Test evaluation is an important higher-order function. Test scripting does not appear to have been investigated at all, though this may be subsumed within test automation, or alternatively may be understood as a 'low-level' clerical or administrative activity.

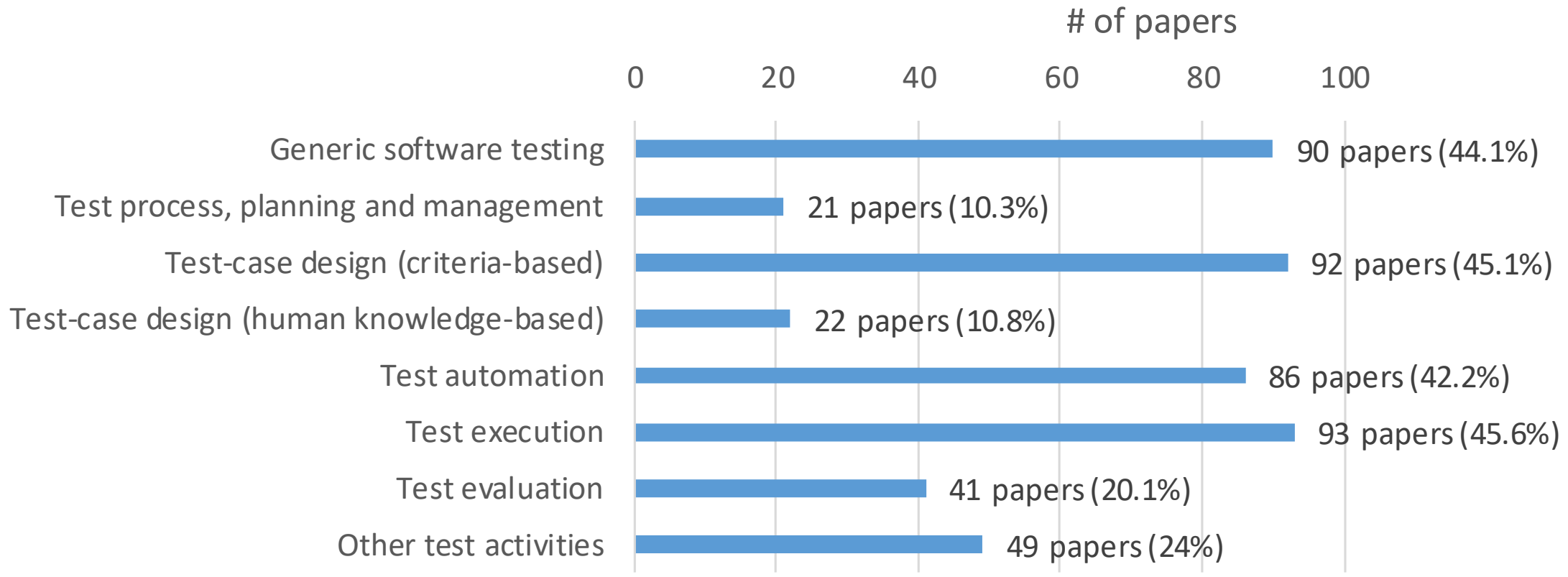

**Figure 11- Frequencies of type of test activities covered in the courses**

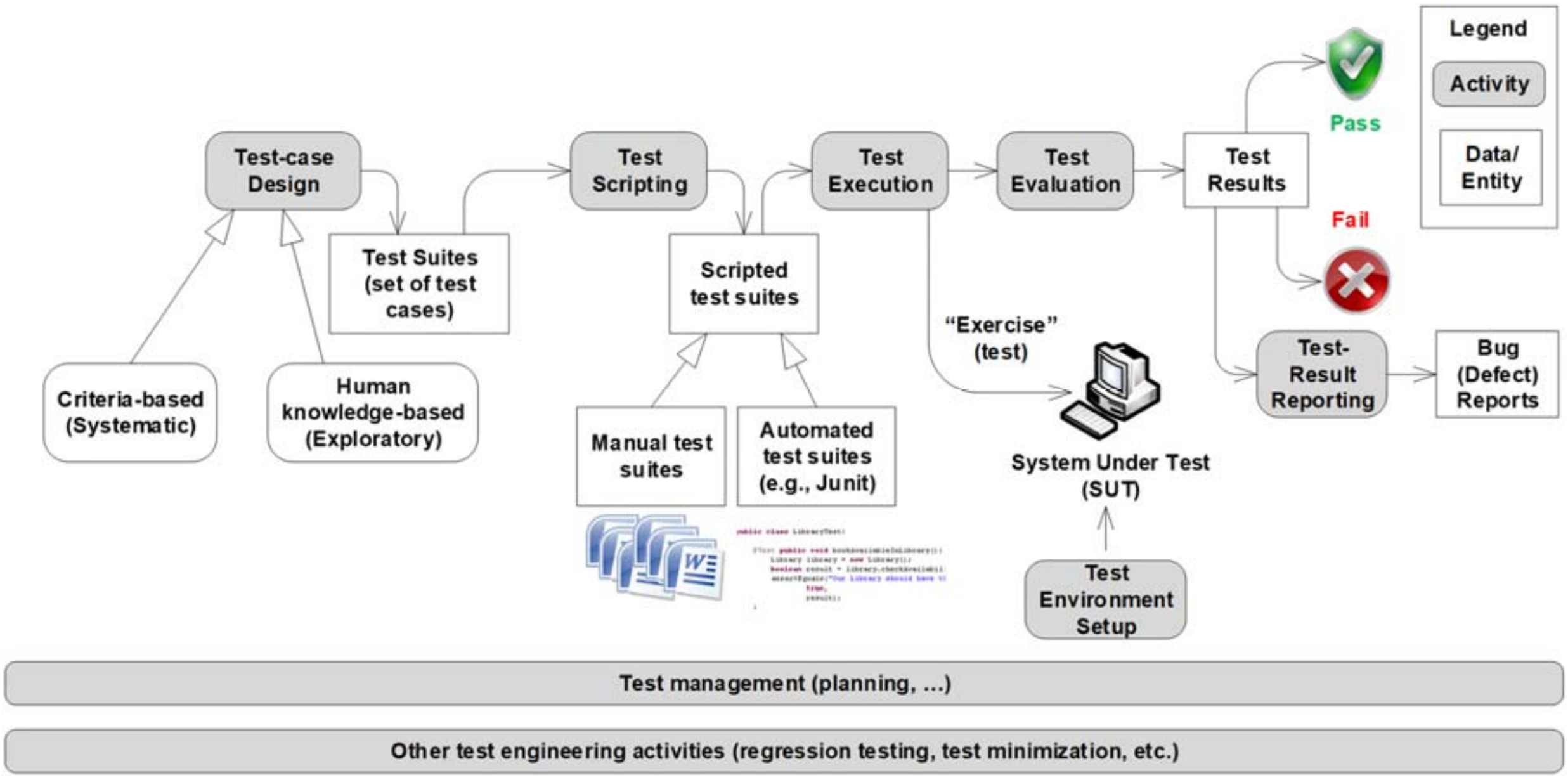

**Figure 12- Overview of a typical software test process and its activities (from [92])**

### 4.6 RQ 6- SCALE OF THE EDUCATIONAL SETTING UNDER STUDY

115 papers mentioned the number of course instances that formed the basis of the published paper. We summarize those data as a histogram in Figure 13. As we can see, in half the cases (59 papers, 51%), the papers discussed only one single instance of a testing course. The paper with the most instances was [P65], in which the evaluation included "*data collected over five years (10 semesters) from 49,980 programming assignment submissions by 883 different students*'. This is an impressive dataset. The average of the number of instances of courses reported across the 115 paper was just under two course instances (1.92).

As with RQ 5, we acknowledge that the papers in our set of primary studies are a (very) small sub-set of all testing courses taught at universities world-wide. Thus, like RQ 5, RQ 6 is also not aiming to provide a world-wide view on the scale of the educational settings in "all" testing courses world-wide, but rather only in the set of the primary studies under review in this work.

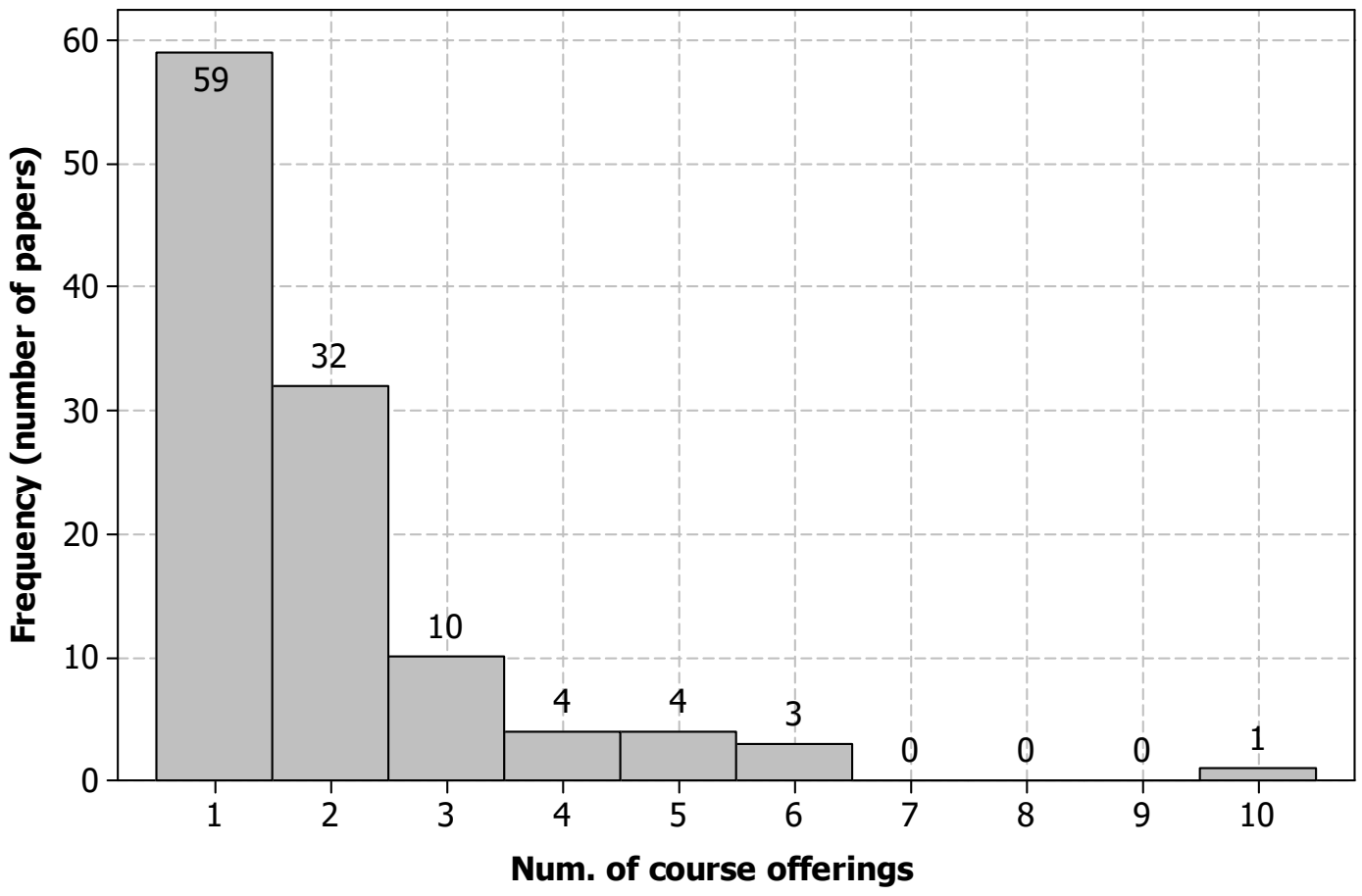

**Figure 13- Histogram of the number of course offerings (n=115 papers)**

102 papers report the number of students enrolled in the testing courses discussed in each paper. Figure 14 summarizes the number of students per paper as boxplots. The number reported in one paper [P2] is clearly an outlier: it reported that, "*nearly 4,000 students from more than 300 universities in China were enrolled*" in the course. Thus, we have visualized the boxplot with and without the outlier in Figure 14.

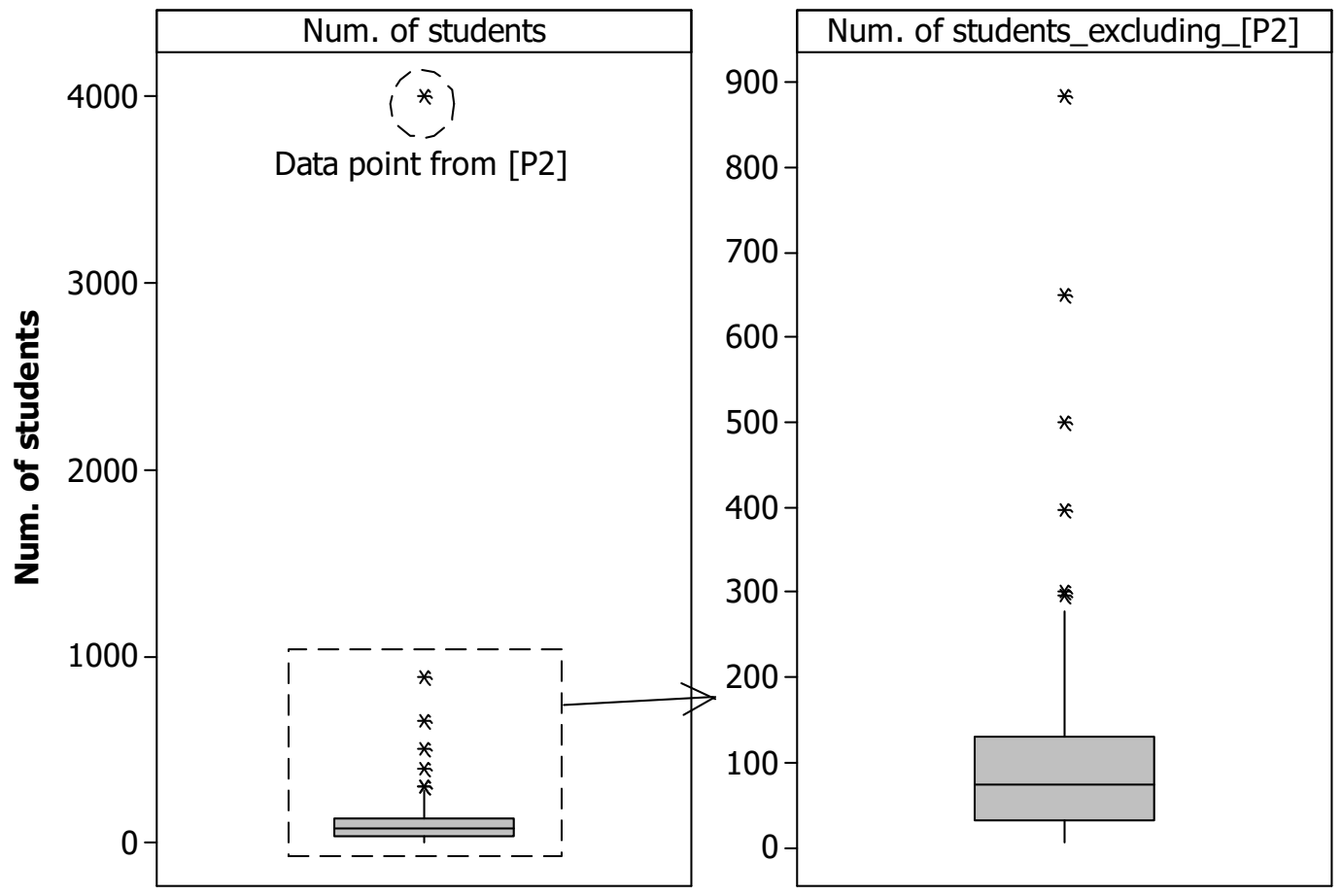

**Figure 14- Boxplots of the number of students enrolled in testing courses (n=102 papers). Right: The same data while excluding the outlier, [P2].**

### 4.7 RQ 7- DIFFERENT APPROACHES TO TESTING EDUCATION: OFFERING SEPARATE TESTING COURSES OR INTEGRATING TESTING IN OTHER COURSES

We looked at whether software testing was taught as a distinct course, separate from other courses, or whether software testing was integrated in some way with other courses. Our classification is shown in Figure 15.

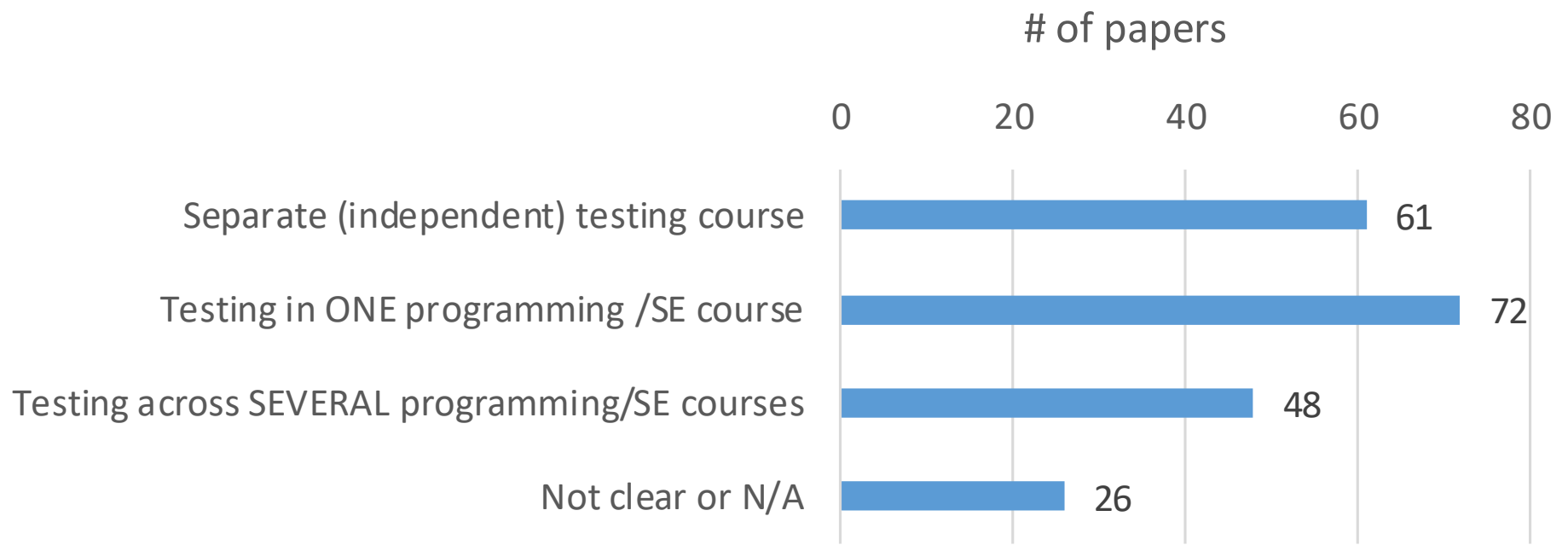

**Figure 15- Frequencies of papers based on how teaching is taught (single testing course or testing across courses/program)**

As can be seen in Figure 15, there is a relative balance among the different approaches to testing education. In nearly one third of the papers, there was a dedicated course on software testing. However, there were more cases in which testing was taught as part of a regular programming course. Both approaches have advantages and disadvantages. On the one hand, a dedicated course on software testing would allow more time and resources to cover this complex subject in more detail. On the other hand, software testing is often considered "tedious" (see RQ 8), and a full dedicated course might face challenges in motivating students, e.g., students lack the opportunity to appreciate the value of software testing. Furthermore, what is learned in a single course might be easily forgotten, if it is then not used nor required in any following course. At times, it might simply not be practical to have a separate course, or alternatively an integrated course, due to time constrains in the curriculum, e.g., "*It is not practical to offer a separate course in software testing, so relevant test experiences need to be given throughout core courses*" [P10]. And even if it was viable to have a dedicated course on software testing, several authors argued that just a single testing course is not enough, e.g., "*Our conclusion is that a separate course in software testing should not be the only place to incorporate testing into the curriculum*" [P138].

A possible solution for these problems is to spread the teaching of software testing across several programming courses, starting already in the early programming courses, e.g., "*there is a need to introduce testing early in the sequence of programming courses, and integrate and continue reinforcing it across all programming courses, rather than delegating it to a single course*" [P154]. Software testing should become a required common task throughout the whole curriculum, e.g., "*the approach must be systematically applied across the curriculum in a way that makes it an inherent part of the programming activities in which students participate*" [P130]. But this also provides several challenges, e.g., "*it is not immediately apparent to students and instructors how to best use tools like JUnit and how to integrate testing across a computer science curriculum*" [P202] and "*Several educators and researchers have investigated innovative approaches that integrate testing into programming and software engineering (SE) courses with some success. The main problems are getting other educators to adopt their approaches and ensuring students continue to use the techniques they learned in previous courses*" [P203].

### 4.8 RQ 8-Theories and theory use in software-testing education

Previous research [58-60] has argued for the importance of using and adapting theories from CS and SE education to increase research rigor. In their book, for example, Fincher and Petre [58] argue that papers in CS education research can be understood to have two dimensions: argumentation (or theory) and empirical evidence. Fincher and Petre [58] develop a diagrammatic representation, based on Pasteur's quadrant [93], of the relationship between theory and evidence. A version of this diagram is reproduced in Figure 16.

In the top-left of Figure 17, there are papers that contain a lot of arguments, but little or no empirical evidence. For the bottom-left quadrant, Fincher and Petre [59] argue that the quadrant should be empty: ideally papers with no evidence and no argument should not be published. The bottom-right quadrant denotes the papers that are mainly based on evidence (experience), but are rather weak on argumentation or "theory". These types of papers are descriptive and mostly experience-based. Fincher and Petre [59] argue that experience papers are the most common types of papers in the CS education literature. Finally, the top-right quadrant represents papers that consider both theory and evidence. Fincher and Petre [59] argue that most CS education research papers *should* belong to this quadrant (but don't).

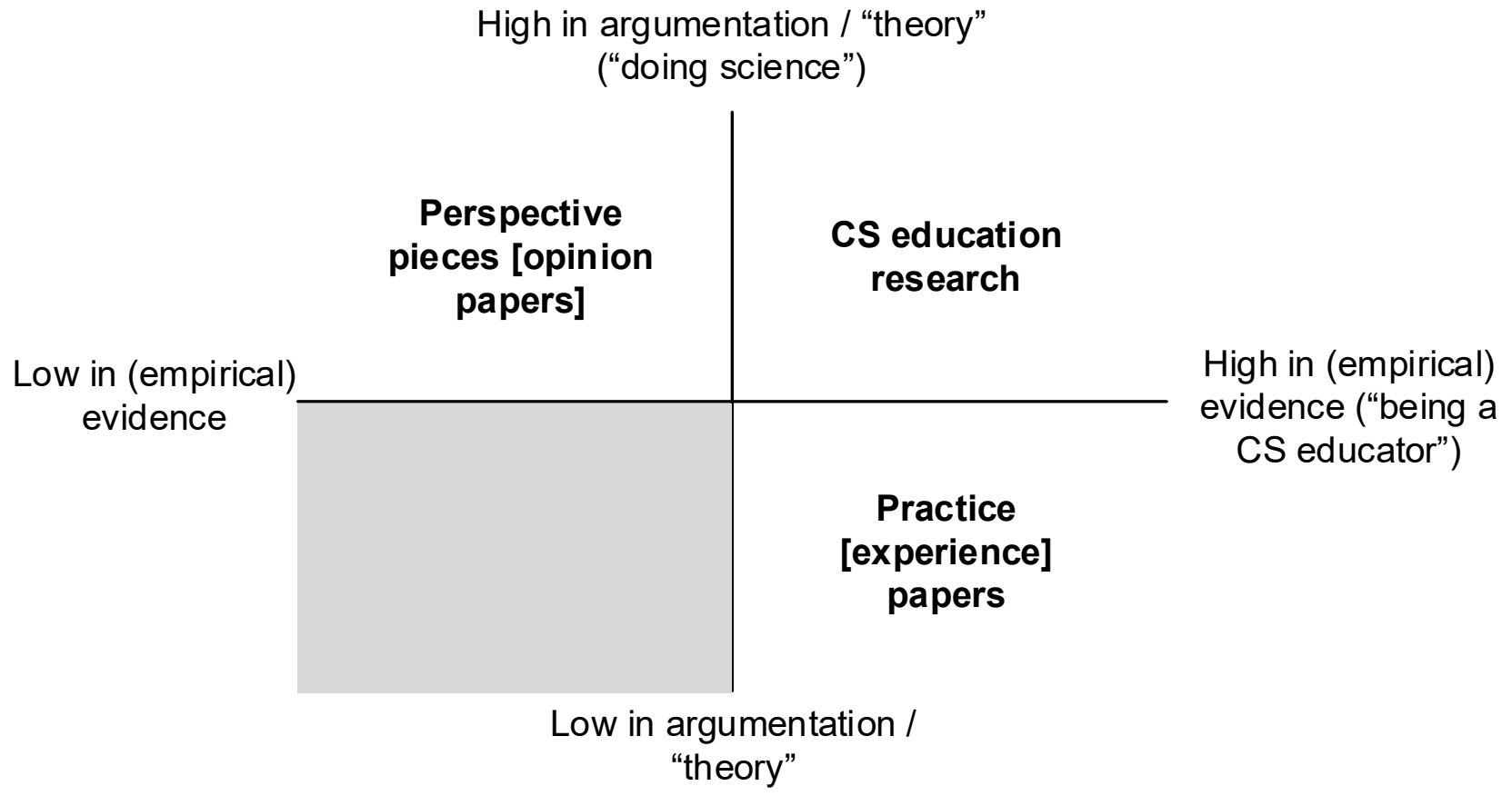

**Figure 16- Diagrammatic representation of the types of papers found in CS education (based on Pasteur's quadrant) (as presented in** [59])

As with Fincher and Petre [59], our RQ 8 explores the state of theories and theory-use in software-testing education. During our process of data extraction from papers, we recorded any paper that used a theory from learning and education science. To clarify, we did not consider "technical" theories of SE, e.g., software-testing theory. Neither did we consider theoretical concepts such as graph theory used in testing. We focused only on theory-use related to the educational aspects of testing.

Surprisingly, only eight of the 204 papers in the pool (3.9%) had used theories. We list those instances of theory-use in Table 8. Popular educational science theories such as constructive alignment theory [94] have been used in a few papers.

The situation in software-testing education literature is therefore similar to the broad literature of CS education, according to arguments of Fincher and Petre's book [59]. Also, similar to Fincher and Petre's recommendations of [59], we argue for more theory use in future papers in software-testing education.

**Table 8: Theories used in the papers in the review pool**

| Paper ID | Theories used | Contexts of theory use |
|---|---|---|
| [P8] | **Cognitive load theory** [95].: This theory offers a method of constructing course pedagogy to help students learn effectively by using an awareness of the limited amount of information that working memory can hold at one time | This paper presents an initial investigation of the impact of integrating the software testing principles with fundamental programming courses by applying cognitive load theory. |
| [P17] | **Constructive alignment theory** [94]: Constructive alignment is a principle used for devising teaching and learning activities, and assessment tasks, that directly address the intended learning outcomes in a way not typically achieved in traditional lectures, tutorial classes and examinations. There are two basic concepts behind constructive alignment: (1) Learners construct meaning from what they do to learn; and (2) The teacher makes a deliberate alignment between the planned learning activities and the learning outcomes. | The course utilized constructive alignment theory by demonstrating continuously the concrete expectation for both the final product (code) and its development process thus allowing students to adapt to match these over the course of the semester. |
| [P20] | **Theory of the zone of proximal development** [96]: The distance between the actual developmental level as determined by independent problem solving and the level of potential development as determined through problem solving under adult guidance or in collaboration with more capable peers. | The challenge of testing education w.r.t. Skill Level and Task challenges was explored by the application of the appropriate teaching approaches to maintain the learners within the zone of proximal development. |
| [P30] | **Constructivist learning theory** [97]: This theory is based on the belief that learning occurs as learners are actively involved in a process of meaning and knowledge construction. | The paper presented an experimental card game for software testing, based on the constructivist learning theory. |
| [P67] | **Diffusion of Innovations theory** [98]**:** It describes how innovations, tools, ideas, or practices perceived as new are adopted by individuals and organizations, and how they diffuse in social systems. | Testing can be regarded as one such idea or practice. Testing is often not taught alongside programming, thus adopting testing practices requires a change in behavior for no immediate or guaranteed benefit. The question, then, is: how can we expose students, novices, or junior developers to experiences that help them adopt testing practices? |
| [P92] | **Constructivist learning theory**: Definition was provided above.<br><br>**Bruner's discovery learning theory**: Discovery Learning is a method of inquiry-based instruction, discovery learning believes that it is best for learners to discover facts and relationships for themselves [99].<br><br>**CDIO educational theory (Conceive, Design, Implement, Operate)**: The CDIO Initiative is an educational framework that stresses engineering fundamentals set in the context of conceiving, designing, implementing and operating real-world systems and products [100]. | Implemented a software testing course based on those three theories. |
| [P94] | **Theory of cooperative learning**: Cooperative learning is an educational approach which aims to organize classroom activities into academic and social learning experiences. It is based on social interdependence theory [101]. | It is important to ensure that all members of student groups actively participate in doing the practical exercise. To achieve this goal, the theory of cooperative learning can be explored. Scholars agree that cooperative group learning when used as a teaching strategy improves problem-solving skills of students. An essential element of cooperative learning is positive interdependence. In order to achieve positive interdependence it is suggested to structure the practical exercises as group exercises such that each member of the group has a specified role and all roles will be linked together. |
| [P162] | **CDIO theory**: Definition was provided above [100]. | The authors developed a new method for teaching software testing based on CDIO, which has these basic characteristics: a project is the main focus, with the teacher as the guide and students are the target. |

### 4.9 RQ 9-EMPIRICAL EVIDENCE / FINDINGS

We organize RQ 9 into RQ 9.1 and RQ 9.2, and discuss the two sub-questions in the following subsections.

#### 4.9.1 RQ 9.1- Challenges in testing education

As discussed in Section 3.6, we extracted the discussions about challenges when teaching testing, as reported in the respective papers. We then used qualitative coding to synthesize the qualitative data. 82 of the 204 papers (40.1%) explicitly presented some form of challenges. We identified 123 text phrases from those papers. As discussed in Section 3.6, in synthesizing and reporting challenges, there were cases where we did not necessarily agree that the reported "challenge" is actually a challenge. For example, [P116] reports that: “*Many of the students found JUnit to be too complicated for them*”. In all cases, we report the challenges as they have been presented in the source paper, and we report these challenges without advocating the efficacy of the challenge. We present in Figure 17 the list of challenges, in which we have organized them into several categories, e.g., those that relate to instructors, instructors and students, and course-design. We discuss each of the challenge categories below and provide a few example papers, which have discussed those challenges.

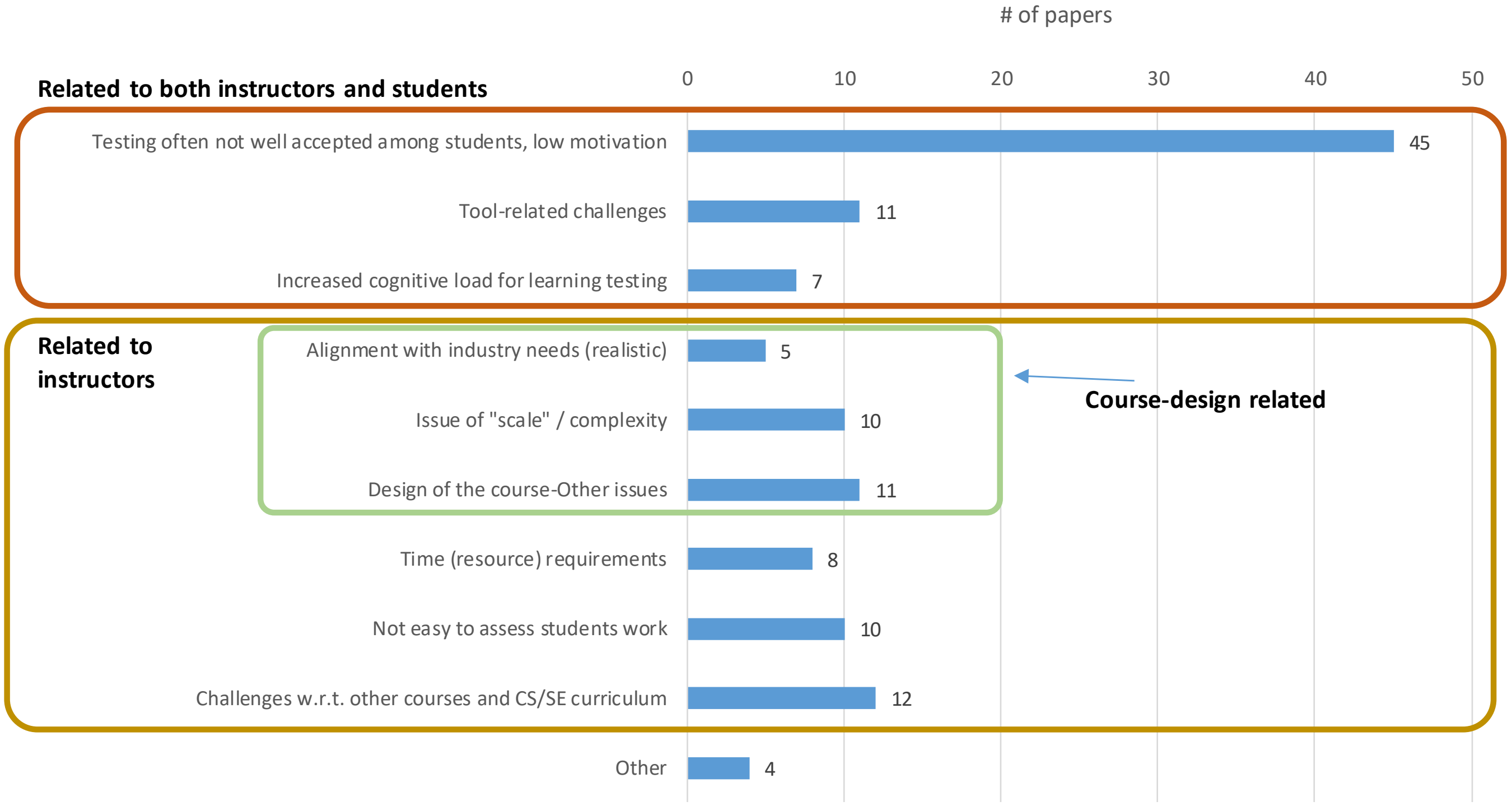

**Figure 17- Challenges in testing education, as presented in the papers**

When extracting qualitative and descriptive data from the papers about the challenges for testing education, it was important to pay a close attention to the "level" of empirical evidence which a given paper had used to extract and report the corresponding challenges. It was important that the reported challenges were indeed based on empirical evidence. We had already classified each paper in terms of the type of research method used, according to the following three categories, (see Section 4.2): (1) Proposals of ideas or approaches for testing education, with no explicitly-mentioned experience in the paper, (2) Experience / informal evaluation, and (3) Experiment / empirical study. We thus analyzed that data for the subset of 82 of all the 204 papers, which had reported challenges. We show in Figure 18 the barchart of that subset of papers according to the above three levels of research method.

47 of the 82 challenge-reporting papers had "experience"-based evidence to support the challenges that they had reported. For example, the authors of [P18] mentioned that: "*the most difficult challenge incorporating unit testing in an experiential course was ensuring students over-come their negative bias to discover the benefits of functional testing*". The authors had come to that observation based on their experience of including unit testing exercises in a university course. As discussed in Section 4.2, we designed the categorization of the above three research method types in a way to be able to interpret them as three increasing "levels" of evidence: (1) Proposals with no explicitly-mentioned experience, (2) experience, and (3) empirical study. Studies which reported empirical studies can generally be seen as having the highest (most rigorous) level of evidence. 29 of the 82 challenge-reporting papers synthesized the challenges of testing education, based on the empirical studies that they had designed and conducted.

Only 6 of the 82 papers reporting challenges did not contain explicit empirical evidence. Similar to the discussion in Section 4.2, we observed that although the authors had not shared any experience of applying the ideas / approaches in their testing courses, it was clear that these authors are active testing educators, and thus at least implicitly, they had tried or were going to try the ideas / approaches in their testing courses.

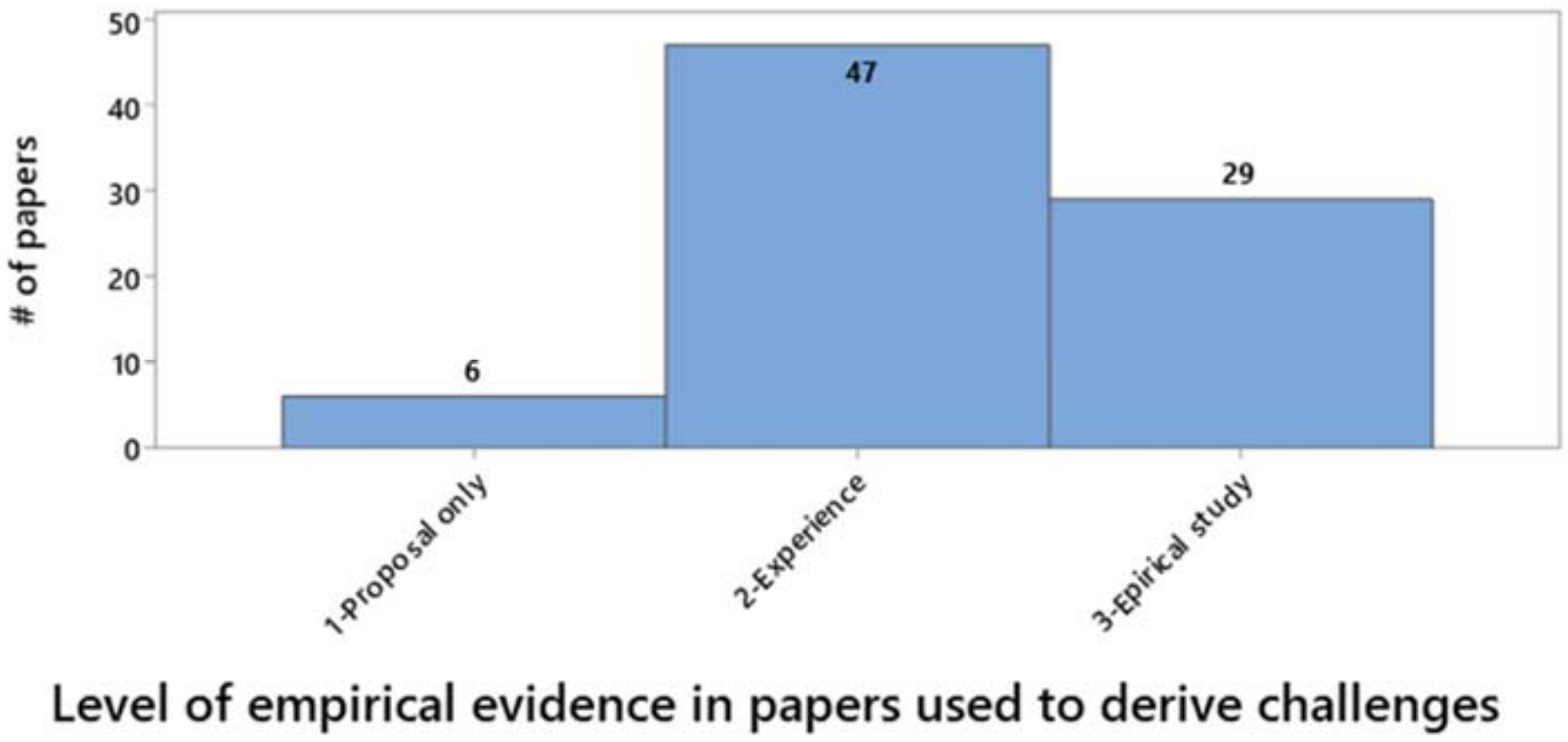

**Figure 18- Barchart of the papers, according to the level of empirical evidence for reporting challenges in testing education**

#### 4.9.1.1 Challenges related to both instructors and students

Testing often not well accepted among students, low motivation:

By far, the most common challenge in teaching software testing is that students do not like learning about software testing. In students' responses to the surveys reported in the papers we reviewed, it is not uncommon to see software testing described as "tedious" and "boring". Students taking a degree in software engineering, or related disciplines, are often much more interested in developing software, and less interested in the systematic testing of what they have developed, e.g., "*While students often derive a great deal of satisfaction from seeing a program that they wrote solve a problem, they do not derive that same sense of satisfaction from exposing flaws in their own programs*" [P23]. And [P20] observes, "*A major challenge was to dispel the stigma of software testing and maintenance as an unholy alliance of arguably the two least favored tasks within the realm of the software life cycle*". [P44] states, "*Software engineering students typically dislike testing… testing can be seen as tangential to what really matters: developing and documenting a design addressing the requirements, and constructing a system in conforming to the design*".

[P183] observed a more general aim in software testing education: "*The challenge for the lecturer is to instill a desire in the students to learn more about the topic and encourage the practice of this specific discipline once they get into industry*".

Tool-related challenges:

To do (automated) testing, students need to learn new tools and libraries, which are often not easy to learn, especially in the earlier (first or second) years of a degree. Using testing tools is a challenge for beginners, especially when they might still struggle with basic programming concepts, e.g., "*Often times, students become overwhelmed with the software testing tools they need to learn to conduct automated testing*" [P165]. Software testing, and especially its automation, requires a good knowledge of programming: "*Testing methods are impossible to understand by students without programming experience, and their depth of programming practice is also a factor*" [P75] and "*testing is a programming intensive activity. The students whose programming skills are rusty have a difficult time with the programming aspects of testing*" [P104]. Furthermore, although nowadays most languages have good support for *unit* testing, more complex testing often lacks good, easy-to-use tools, e.g., "*Unfortunately, students and instructors continue to be frustrated by the lack of support provided when selecting appropriate testing tools*" [P196].

Increased cognitive load for learning testing:

Another factor affecting students' attitudes toward software testing may be the increased cognitive load placed on learning about testing. [P21] stated: "*This [TDD] does require a change in thinking and does not come naturally to all students.*". [P51] made a similar observation, "*Research has noticed that imparting TDD-like testing to an early computing curriculum is challenging because it increases technical and cognitive load for the students.*" [P119] reflected that: "*Learning to program is hard. Why make it harder, by requiring students to learn additional syntax in order to express their test cases?*"

[P123] argued that teaching software testing skills for first year students in CS courses can be particularly challenging, since besides learning the programming basic structures, students have to deal with peculiarities of the specific techniques and tools for software testing.

**4.9.1.2 Challenges related to instructors:**

Challenges related to course-design:

- Alignment with industrial skill-set needs (theory vs. practice): There is recognition of the value of making software testing courses practical and close to industrial needs, but in doing so there is also the ongoing challenge of keeping the courses that "remain" aligned with industry. [P125] recognized this challenge, stating that "*From the educator's perspective, it is hard to keep a testing course up-to-date with the novelties of the field as well as to come up with exercises that are realistic*". Also, it has been reported that testing concepts in some university courses are only taught theoretically, with no or little practical experience, e.g., "*University courses do not seem to provide practical knowledge or experience, and students do not develop a habit of testing… Although testing is taught in courses, students have no practical experiences in testing and even so, students may have forgotten how to test correctly*" [P67] and "*However, they complained about software-testing education being too much theoretical, with a lack of practical scenarios to show students how the concepts should be applied and how software testing would have an impact in the medium and long term*" [P15]. This approach simply does not work, as testing is intrinsically a practical activity, e.g., "*Disconnection between theory and practice leads to less interest by students*" [P162]. But teaching the practice of testing faces the issue of finding the right case studies, which should be of enough complexity to warrant the need for testing, but not so complex as to overwhelm the students. Using real software projects as case studies would help, but finding (e.g., on open-source repositories) or implementing (by the instructor) the right projects to use in software-testing education is not trivial.
- Issue of "scale" / complexity: One aspect of realism is scale: testing large-scale software systems, and conducting large scale tests of a software system. [P53] states, "*Students perceived test writing to be irrelevant due to the small size of the programming tasks*". [P147] stated: "*The problem is that rigorous testing is more costly than beneficial in the small-scale projects and exercises that are usually given in software engineering courses*". Another paper, [P146], stated that: "*… developing software tests for programs that have significant graphical user interfaces is beyond the abilities of typical students (and, for that matter, many educators)*". When only dealing with the coding of small programs/functions, students might not fully understand the benefits and necessity of software testing, e.g., "*Students perceived test writing to be irrelevant due to the small size of the programming tasks*" [P53] and "*The problem is that rigorous testing is more costly than beneficial in the small-scale projects and exercises that are usually given in software engineering courses*" [P147].
- Other issues related to course design: [P100] observed that: "*Learners need constant and concrete feedback on how to improve their performance on testing at many points throughout the development of a solution rather than just once at the end of an assignment*". [P108] recognized the need to respond differently for students with differing abilities: "*Weaker students need continuous feedback that they're on the right track, while stronger ones prefer freedom.*"

Time and resource constraints:

A number of papers recognized the time and resource constraints that impact the design, delivery and assessment of software, for example: "*[There is] not enough time to teach testing in programming courses*" [P9], "*Many studies have cited limited time (both preparation time and instruction time) as one of the major barriers to teaching software testing*" [P23], "*It is challenging to evaluate thousands of assignments within limited time*" [P50] and "*The overriding challenge is that there are usually too many topics to be covered in [software] courses and there is little or not time to teach testing*" [P103].

Challenges related to assessing students' work:

Authors recognized challenges in assessing students work on software testing. [P11] recognised the negative side of gamification: students 'game' the assessment system. [P16] expressed concerns around the opportunity for instructors to assess the quality of the software testing and not just the tested correctness of a program. The [P16] mentioned that: "*Care must be taken to avoid an observed tendency to approach assignments in a tick list fashion*". [P40] recognised that current assessment techniques based on automated grading tools for evaluating student-written software tests are imperfect. [P108] observes that grading exercises requires a lot of effort, effort that should instead be used on supporting the learning process.

A common approach is to check the code coverage of the implemented tests. But code coverage alone is not a satisfactory measure for test quality. For example, a student could write tests with no assertions on the expected outputs, and such a major issue would not be detected by code coverage alone. Advanced testing techniques like *mutation testing* [82] could help in such regard, but, unfortunately, it is not so well known outside of test specialists, and tool support is still rather unsatisfactory (and both reasons could explain why mutation testing is still not so widespread). There is a dire need to be able to evaluate the quality of the test cases *automatically*. Educators have often limited time to evaluate students' assignments. Also having to evaluate the test cases manually would be often not viable, e.g., "*It is challenging to evaluate thousands of assignments within limited time*" [P50].

Challenges related to integrating software testing in other courses:

Papers also reported a number of challenges related to integrating software testing in other courses. [P10] mentioned that in their university program, there was no compulsory testing course and, given program constraints, it was not practical to offer a separate course in software testing.

[P114] argued that "*any discussion of how best to teach testing at the undergraduate level is complicated because there are several different types of software-related undergraduate programs, all with differing goals and priorities*". For example, universities offer programs in CS, computer engineering, SE, and information science, each emphasizing different aspects of software, and each having different amounts of time available to devote to testing and other SE issues. Furthermore, it mentioned that: "*what to include in other courses and how much to split off into V&V specific courses is a difficult question in curriculum design*". Authors of [P142] also found it challenging to determine "*the optimal progression through a SE course structure in regard to software testing*". [P191] reported that a critical issue for the success of the integrated teaching of testing and programming foundations is how to provide appropriate feedback and how to evaluate the student's performance.

#### 4.9.1.3 Other challenges

There were other reported challenges, which could not be classified under the above challenge categories. Eight papers reported such challenges. For example, [P21] mentioned that: "*textbooks do not cover test automation [very well]*". [P75] stated that, "*Unfortunately, no definitive 'Theory of Software Engineering' exists unlike the theoretical concepts that underpin other fields in computer science that are more amenable to mathematical descriptors*". And [P161] provided an interesting point by mentioning that it is difficult to interest student programmers in thoroughly testing their own programs since "*every fault found represents a psychological blow to their programming ego*".

### 4.9.2 RQ 9.2- Insights, observations, and recommendations for testing education

Similar to our analysis of the challenges in testing education, we extracted the insights, observations and/or recommendations presented in the papers for effective teaching of testing. We found that many important and interesting insights, observations and/or recommendations for testing education were provided in the papers, which were based on experience and/or evidence. As defined in the English dictionary, an insight is "*the capacity to gain an accurate and deep understanding of something or someone*". Many papers report such understandings.

Similar to the discussions in Section 4.9.1, when extracting qualitative and descriptive data from the papers about the insights for testing education, it was again important to consider the level of empirical evidence which a given paper had used to extract and present the reported insights. We used the same classifications as used above: (1) Proposals of ideas or approaches, with no explicitly-mentioned experience in the paper, (2) Experience, and (3) Empirical study. We analyzed the data for the subset of 152 of all the 204 papers, which had reported insights. We show in Figure 19 the barchart of that subset of papers according to the above three levels of empirical evidence. As we can see, once again, a large proportion of insight -reporting papers have used either "experience"-based (79 papers) or empirical-study-based evidence (67 papers) to derive and support the insights that they have reported. For example, the authors of [P3] conducted an empirical study, by analyzing students' programming projects and also by interviewing students at the end of the academic term. From the findings, paper [P3] identified potential for influencing student testing behaviors. Based on empirical study results, [P3] presented the following insights that: "*The students who expressed particularly strong reticence to follow Test-Driven Development (TDD) also happened to be students who were especially confident in their programming skills. The proximity of testing and its frequent association with debugging suggest that within students' mental models, testing is a process for fixing problems rather than proactively avoiding them. In order for students to adopt different behaviors, this mental model would have to change*".

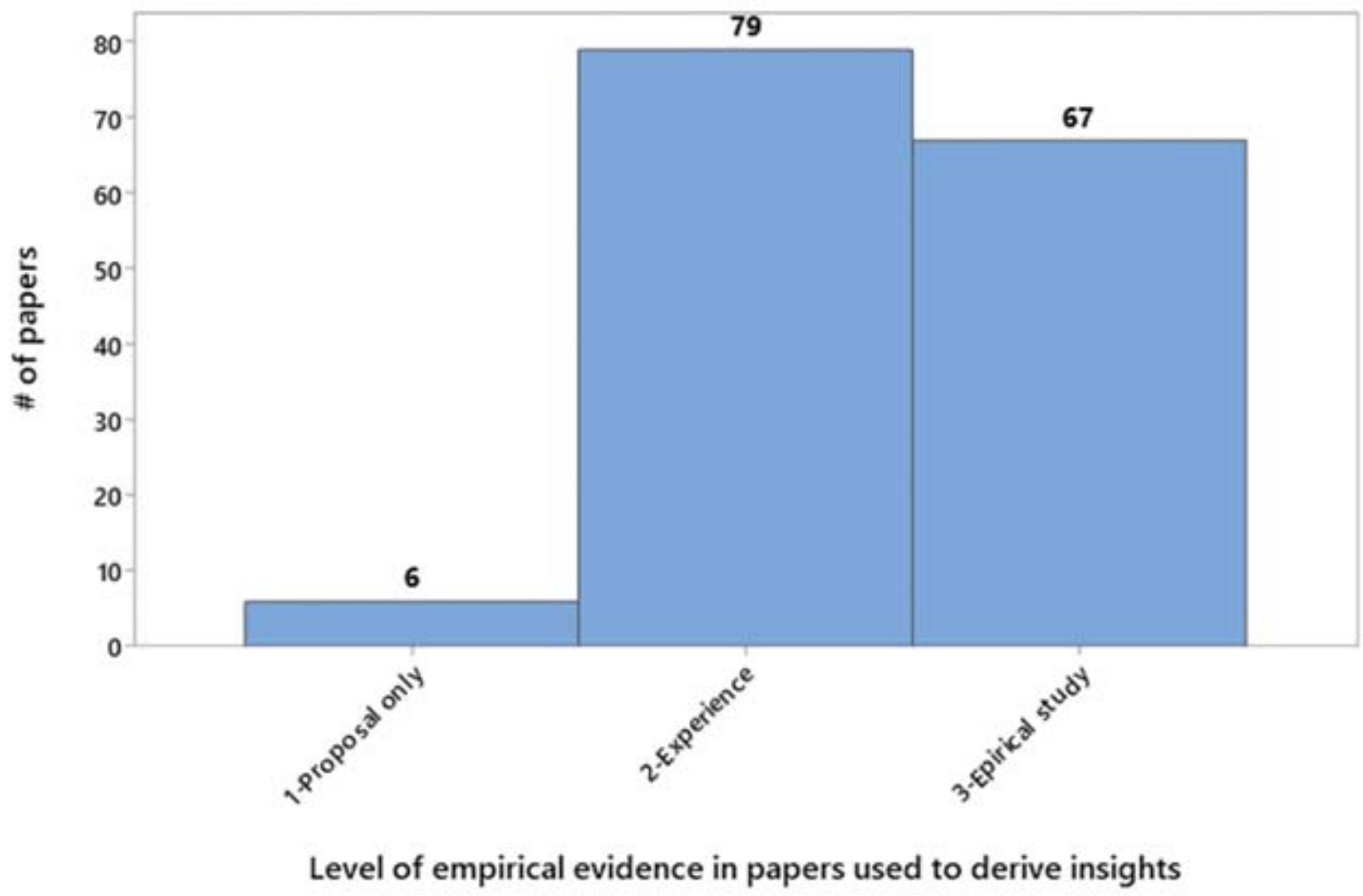

**Figure 19- Barchart of the papers, according to the level of empirical evidence for deriving insights for testing education**

We then synthesized the extracted data about insights, observations and recommendations using qualitative coding (as discussed in Section 3.6). 152 of the 204 papers (74.5%) presented some form of insights, observations and/or recommendations. This is a much greater percentage than papers reporting challenges (~40%, as discussed in Section 4.9.1). 233 text phrases were identified from that subset of 152 papers. Figure 20 presents the results of our qualitative coding of insights.

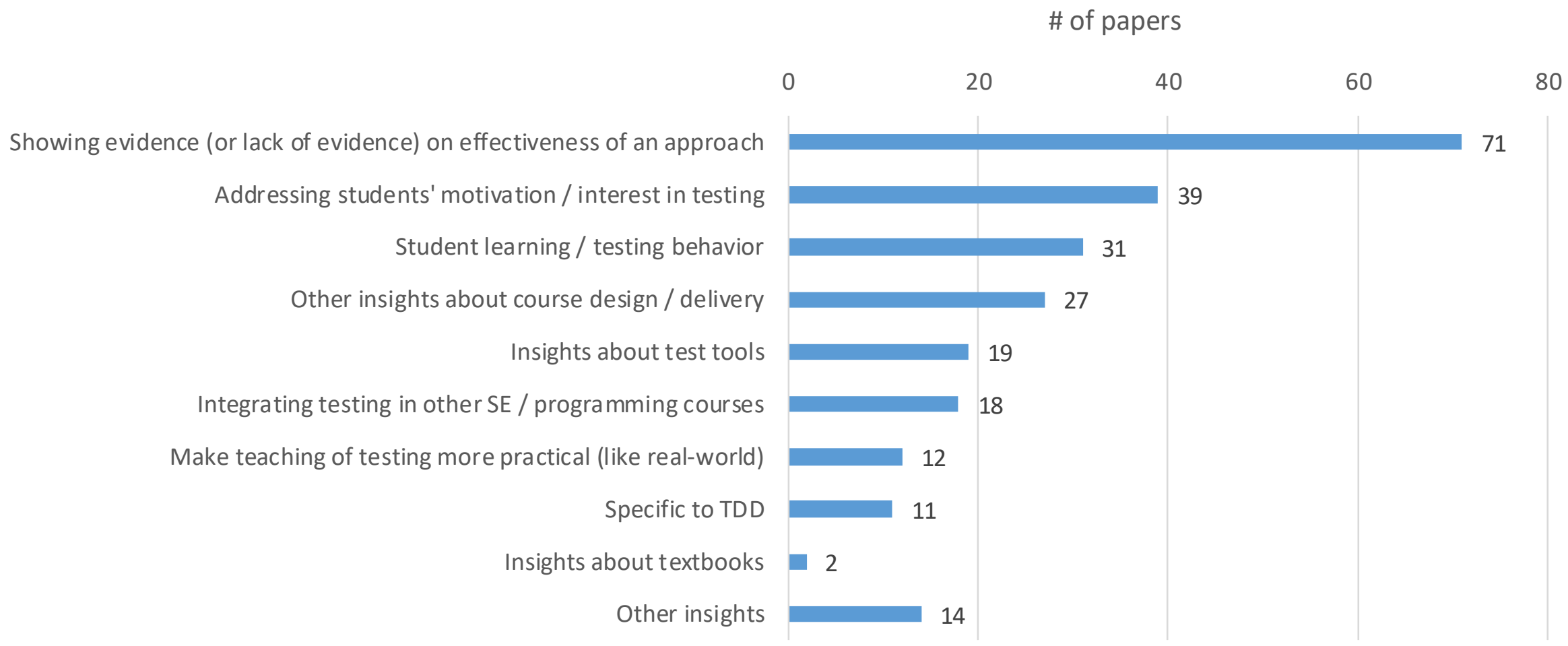

**Figure 20- Insights, observations and recommendations for testing education, as presented in the papers**

The surveyed papers presented many types of insights into the teaching of software testing, especially regarding how to address the previously-discussed challenges. The most common category of insights, observations and recommendations was those related to the introduction of a teaching approach or technique, supported by evidence to claim its effectiveness at improving the learning outcome of the students. We briefly review next each of the categories, shown in Figure 20, and provide a few example papers in each category.

**Insights showing the evidence on effectiveness of presented teaching approaches:**

The most common case was the introduction of a teaching approach or technique, supported by evidence to claim its effectiveness at improving the learning outcome of the students. For example, [P28] reported that: *"The results provide evidence that the reuse of test cases during introductory programming courses may help to increase the quality of the programs generated by students, motivating them to apply software testing during the development of the programs"*. Another paper, [P46], reported that: "*We found evidence that students who learn to write good defensive programs can write effective attack programs, but the converse is not true… our results indicate that a greater pedagogical emphasis on defensive security may benefit students more than one that emphasizes offense*", and [P176] reported: "*The results of our empirical study provide evidence in favor of greater formal training in software testing as part of CS programs*".

**Insights related to addressing students' motivation / interest in testing:**

As noted earlier, many students find testing to be "tedious" and "boring". Students' attitudes toward testing can also be influenced by their attitudes to university assessment. For example, [P20] reported: "*We found that students placed a strong emphasis on extrinsic motivation such as grades*".

Many educators report on the initiatives they have implemented to seek to change students' attitudes to software testing, and to improve the experience of testing and the learning about testing. *Gamification* seems to be a promising initiative to improve the experience of (learning about) testing, by trying to make testing like a "fun" activity, e.g., papers [P5,P30,P48]. As students may not have the appropriate mindset for testing their own software, different educators report benefits of having them test software written by their peers. For example, [P18] reported that: "*It is often easier to convince a programmer to test someone else's code rather than try to convince them that their code requires testing*", [P23] reported, "*When testing their own code, students are less motivated to find bugs, as bugs expose their own failure to develop a correct program*", and [P199] reported, "*Peer review, or peer testing, in which students attempt to break code written by their peers, has the potential to address all of these problems. Because it is competitive, it can be fun and exciting*". Using real-world software (e.g., [P50,P74,P195]), or even just guest lectures from industry [P13], can be very beneficial to make the students more motivated. Just requiring students to do testing as part of their assignment is not enough: "*Students need to directly experience benefits from writing test suites. Requiring students to write test cases simply because test suite quality will be graded does not help students learn the value of testing*" [P85].

**Insights related to student learning / testing behavior:**

Related to student motivation is student learning and testing behavior. Changing students' mental models of the purpose of software testing can help to motivate students, and also help students to adapt their behavior both for learning about software testing and then applying that learning into real situations. For example, [P3] reported: "*The proximity of testing and its frequent association with debugging suggest that within students' mental models, testing is a process for fixing problems rather than proactively avoiding them. In order for students to adopt different behaviours, this mental model would have to change*", [P16] reported: "*… software testing tends to move students towards a reflective approach to programming, and away from a trial and error approach*", and [P124] provided a cautionary note: "*… stellar performance on examinations doesn't mean that students can transfer the knowledge beyond the classroom*". [P125] is an example of an investigation that identifies a range of common testing mistakes made by students, the topics students find hard (or hardest) to learn, and the teaching methods students find most helpful.

**Insights related to course design, delivery and assessment:**

Authors also discussed course design, delivery and assessment. Inevitably, there are connections here with making software testing interesting and meaningful to students, and encouraging students to adopt appropriate mental models and behaviors for effective and efficient software testing. Earlier in this paper (e.g., Figure 9), we identified a proportion of papers that report on proposals for software testing, and one would expect such proposals to consider course design, delivery and assessment at some level of abstraction. [P32] reflected: "*Where do we cut previous content?, and: When, and to what extent, do we cover the theory of software testing in our teaching? Another, more content related question is: How do I know that I've written the right tests, and enough of them?*". [P58] reported that: "*It is imperative to limit the scope and depth of the course. Software testing is too wide a field to be covered in one single course, much less for people who do not have a computer science background.*"

**Insights about test tools:**

A common approach adopted in previous studies is to use a tool, or tools, as a vehicle for supporting the teaching and the assessment of software testing. Section 4.1.2 identifies a number of testing tools that have been used more frequently in previous studies. [P37], made an observation that applies more widely: "*Despite the weaknesses of these testing tools, we found them indispensable. ... would our software engineering courses have been so rich [without them]*". [P84] advised: "*Use popular, widely used testing tools rather than tools designed for education, in order to teach students the correct use and configuration of real environments.*"

**Insights related to integrating testing in other SE / programming courses:**

We have already contrasted, in Section 4.7, the teaching of software testing as a separate course compared to integrating the teaching of software testing into other courses, particularly programming courses. Here we provide some illustrative quotes from papers. [P29] stated: *"Because students tend to compartmentalize knowledge to a single course, and not transfer it to new situations, we felt that an incremental, just-in-time introduction of testing practices would work better than a separate course. The testing activity is inserted in a value-added manner that does not disrupt or compromise course content or flow*". [P52] stated: "*It is clear that there is a need to integrate software-testing education with other disciplines along the CS undergraduate courses.*" [P67] took a different perspective: "*Just as they had been taught programming before, our results suggest that there is a need for purely testing-oriented projects that let students focus on this part of software development… Educators might need to consider providing additional courses solely focused on testing...*"

**Insights on how to make teaching of testing more practical (like real-world):**

Authors appreciated the value of real-world scenarios for software testing, for example to motivate students and to help ensure students gained relevant experience and learned relevant skills. A notable perspective is presented in [P179]: "*The majority of students do not have the level of understanding required by industry to test their systems. This proves that their understanding of the tests used by industry in software testing is not in line with those of industry. Significant differences were found between software testing skills required by industry and those claimed by students*". As another example, [P44] stated that: "*Students need the opportunity to put what they learn into practice, using testing techniques and tools at all levels, from the individual unit to the system as a whole.*"

**Insights specific to using Test-Driven Development (TDD) in testing courses:**

A recurrent topic among the analyzed papers is Test-Driven Development (TDD) [15]. Although its application in industry is not so widespread [102], many educators investigate whether it can be useful for teaching software testing. Like TDD's application in industry, opinions on the use of TDD in education are mixed. On the one hand, some educators reported positive experience with TDD, e.g., "*I believe TDD is useful in education because it provides the student with timely feedback during development and helps them complete assignments using small, focused steps*" [P21], "*Introducing TDD early will provide multiple positive outcomes*" [P9], and "*The results have been extremely positive, with students expressing clear appreciation for the practical benefits of TDD on programming assignments*" [P96].On the other hand, there are many challenges to the use of TDD as a didactic tool, e.g., "[...] *students particularly struggle with TDD because its test-first approach is 'almost like working backwards'*" [P3], "*This [TDD] does require a change in thinking and does not come naturally to all students*" [P21], "*The test-first aspect of TDD garners more resistance than unit testing*" [P80] and "*TDD is not cost-free. It requires knowledge of testing frameworks and skills in their use, an understanding of refactoring, and an unlearning of old habits from test-last development. Rigorous evaluation of the purported benefits of TDD yield mixed results*" [P172]. It appears that TDD can be beneficial for didactic purposes, but educators that want to introduce TDD in their courses must be aware of its challenges, and properly address them.

**Insights about textbooks:**

Only two papers provided hints about textbooks. [P58] mentioned that: "*Pre-class readings from an appropriate textbook facilitate the learning process*", which is a rather common recommendation in education. [P75] raised the "*importance of good textbooks*" in testing courses.

**Other insights:**

In addition to specific insights on software testing, authors also occasionally commented on more general experiences of teaching and educating. [P21] stated: "*Through patience and reinforcement of incremental development using small steps such as continuous refactoring, my experience has been that students eventually get better at it and begin to come to me with small, focused problems instead of bringing me a complete application and asking me: 'Why doesn't this work?'*". [P92] stated that: "*Project-based learning within small groups dramatically improved their teamwork and communication capabilities, as well as development and project management capabilities*". [P197] stated: "*One of the strongest lessons we have learned is that our students need more math background.*"

#### 4.9.3 Relating challenges and insights

To better understand how challenges and insights are related, we analyzed the semantic relationship among challenges and insights presented in the papers and visualize those relationships in Figure 21. Challenges and insights clearly relate to each other, and may affect each other. For example, introducing a new element to a course, such as increasing the realism of the software testing, can lead to an increase in complexity and scale, which can increase the cognitive load on students, require more resources of instructors, and therefore reduce the resources available to effectively assess student learning.

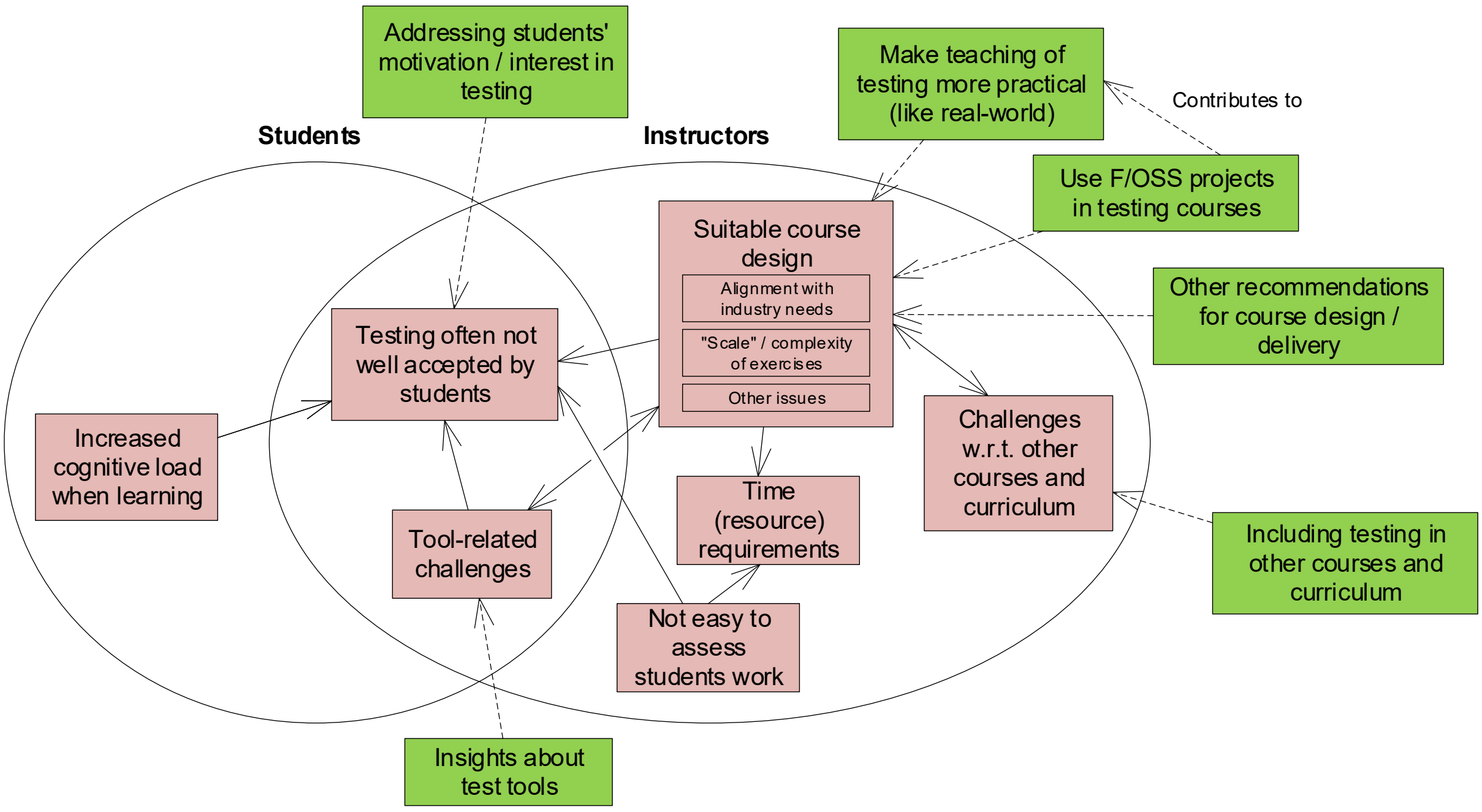

**Figure 21- Relationship among challenges and insights presented in the papers (green-backed rectangles are insights and pink-backed rectangles are challenges.**

## 5 DISCUSSIONS

In this section, we first summarize the research findings and discuss implications and recommendations for educators (Section 5.1). Then, we present suggestions for further education research in this area in Section 5.2, and finally we discuss potential threats to validity of this review (Section 5.3).

### 5.1 RECOMMENDATIONS FOR EDUCATORS

Given the large body of experience and knowledge in the area of software-testing education (204 papers were included in our review), it is often not possible for an educator or researcher to study all the papers and synthesize all the experience and evidence presented in all the papers. For example, for a new educator who wants to teach a (new) course in software testing, it would be valuable to know, before teaching, about the challenges faced by educators when teaching testing and also about insights into how to address those challenges. We recommend new educators of testing to review the list of challenges faced when teaching testing and also consider the insights on how to address those challenge (Section 4.8).

An important decision which has to be made in a given SE-related undergraduate degree program is whether there should be a single testing course or whether testing should be integrated across one or more other (typically programming) courses. As we reviewed in Section 4.7, there are advantages and disadvantages associated with each approach. We therefore recommend that educators and curriculum designers use the suggestions from this SLM, and the primary studies that we have cited in Section 4.7, to help them make an appropriate decision for their context.

The other RQs that we investigated provide suggestions for educators, e.g., RQ 5.1 (type of test activities covered in the courses). For example, we found that most courses have taught criteria-based test-case design, and fewer courses are teaching exploratory test-case design. By contrast, exploratory testing is quite popular in industry [103] and research has also reported benefits with exploratory testing [104]. We thus recommend more coverage of exploratory testing in software-testing education. But to clarify: not all software-testing educators publish papers from their education efforts, and our review paper draws on only a sample of education and is therefore at best a partial “lens” into software-testing education in universities.

As based on our investigation of the RQs in this study, we observed the emergence of the idea for a “design framework” for software testing courses, which would consider all “design” aspects of a given testing course. Such a “design framework” would have multiple dimensions (e.g., choices of the degree of theory versus practice, and choices of pedagogical approach) that will enable educators to systematically design and deliver a given testing course. Similar design frameworks have been proposed in other areas of CS education research, e.g., [105, 106], but we are not aware of any for testing education, in particular. While this SLM provides some insights about such a design task, we recognize the need for such a framework in future works.

As a general comment, each of the challenges and insights provides a kind of 'recommendation', i.e., in the design, delivery and assessment of a given testing course; and we encourage testing educators to consider the synthesized list of challenges and insights in Section 4.9. For example, to address the widely-discussed challenge of "*Testing often not well accepted among students*", testing educators are recommended to seek ways to motivate students to learn about software testing, and to change students' attitudes towards and their expectations of the reasons for, and value of, software testing.

### 5.2 Suggestions for further education research in this area

Research into software-testing education may be understood as design-science [107], where researchers seek a better understanding of the nature of software testing and its education, whilst also seeking to change the way we educate students so that they can become better software testers

Each of the 120-odd RQs presented in Table 6 provides the opportunity for replication or extension to the research. Key research questions, that fall across the spectrum of design science [107], could be:

- How do we motivate students to engage actively with courses on software testing?
- How do we change the mental models of students so that they appreciate the value of software testing?
- What additional measures, or metrics, could be developed to help educators assess the quality of software testing beyond 'just' code coverage?

At a more conceptual level, and as we discussed in Section 4.8, it was surprising to find out that only eight of the 204 papers in the pool (3.9%) had used education theories. This low ratio is quite similar to the broad literature of CS education, as discussed by [59]. Thus, there is the need for more use of education theory in future papers in software-testing education.

### 5.3 Potential threats to validity

This SLM paper has the same kind of threats to validity common to any SLR in the SE literature. In particular, there are potential issues with how data was extracted, limitations in the search terms that might have missed important papers, and bias in the applied exclusion/inclusion criteria. In this section, the threats are discussed using the standard classification in the literature [99].

*Internal validity*: to make sure that the study can be repeated, we used a systematic approach with precise search terms and inclusion/exclusion criteria, as described in Section 3. However, there is a potential threat that relevant articles have been missed. To mitigate and minimize this threat, we used two different search engines (Google Scholar and Scopus) and we snowballed. Therefore, if there are any relevant papers missing from our sample, their rate should be low to negligible. Also, inclusion/exclusion criteria and their application could also be affected by the bias of the researchers, depending on their judgment and experience. To minimize this threat, we used a joint voting and peer reviewing among the authors. Furthermore, as discussed for RQ5 and RQ6, SLMs and SLRS are not primarily a tool for exploring the occurrence of a phenomenon (unless that particular issue has been studied in the primary studies), but for exploring existing knowledge about a phenomenon. Consequently, neither RQ5 or RQ6 aim to provide a globally generalisable view on the type of test activities covered in "all" testing courses, but instead the RQs focus only on the set of the primary studies under review in this work. Thus, the generalizability and implications of the results relating to the study's RQs should be treated carefully. This is an issue common to SLMs and SLRs, and not specific to only this study.

*Conclusion validity*: Conclusion validity of a literature review study is asserted when correct conclusions are reached through rigorous and repeatable treatment. In order to ensure conclusion validity, all related primary studies were selected and all authors reviewed the terminology used in the defined schema to avoid any ambiguity. Data extracted from the primary studies by one author were peer reviewed by another author to mitigate bias. Each disagreement between authors was resolved by consensus among researchers. By following the systematic approach and described procedure, we ensured replicability of this study and strengthened the likelihood that results from similar studies will not have major deviations from our classification decisions.

*External validity*: This study provides a comprehensive review on the field of software-testing education (overall 204 papers are included). Also, note that our findings in this study are within the field of software-testing education. We have no intention to generalize our results beyond this subject area.

## 6 Conclusions and future work

We conducted a systematic literature mapping (SLM) to identify the state-of-the-art in the area of software-testing education. Our SLM provides a classification of studies in this area, a synthesis of both challenges faced during testing education and insights for testing education, and an index to studies in this area. All three contributions can benefit educators in the design, delivery and assessment of software testing courses in university settings, and can provide a foundation of previous research to inform the design and conduct of further research on software testing and software-testing education.

After compiling an initial pool of 307 papers, we then applied a set of inclusion and exclusion criteria to reduce our final pool to 204 papers (published between 1992 and 2019). Our SLM demonstrates that software-testing education is an active and increasing area of research. Many pedagogical approaches (e.g., how to best teach testing), course-ware, and specific tools for testing education have been proposed. Challenges and insights into the teaching of software testing have also been identified and discussed.

We suggest future work in the following directions: (1) Using the findings of this SLM in software testing courses, and evaluating the findings; (2) Comparing the findings from this SLM with the results of previous review studies, as reviewed in Section 2.2; (3) Comparing the state of software-testing education in universities with training in industry (as per the conceptual diagram presented in Figure 1); (4) Using the findings of this SLM for developing a flexible framework to enable software testing educators to "design" their courses based on the evidence and experience in this SLM; (5) assessing the extent to which the results of this SLM (classification and synthesis of data) meet the SLM's needs (as put forward in Section 3.1) by asking the opinion of a sample set of appropriate beneficiaries, e.g., educators of software testing, and also the education researchers in this area.; (6) deriving guidelines from a synthesis of the literature for how the teaching of testing should be conducted; and (7) further synthesizing the evidence presented in the papers that we have reviewed, as we have conducted a preliminary SLM in this paper. Whilst it is encouraging to see the number of experience reports on software-testing education, we also encourage the community to seek to conduct more, and more rigorous, evaluations of the interventions (e.g., tools, courseware, curriculum) used in the courses.

## Acknowledgment

Andrea Arcuri is supported by the Research Council of Norway (project on Evolutionary Enterprise Testing, grant agreement No 274385). We thank the anonymous reviewers and the editor for their constructive comments.

## 7 References

### 7.1 Pool of studies in the systematic literature mapping